\documentclass[letterpaper, 10pt, times, preprint]{elsarticle}

\usepackage[english]{babel}
\usepackage{graphicx,subcaption} 
\usepackage{multirow}
\usepackage[table,xcdraw]{xcolor}
\usepackage{booktabs}
\usepackage{array}
\usepackage{comment}
\usepackage{float}
\usepackage{pgffor} 

\usepackage{float}

\usepackage{graphicx}
\usepackage[percent]{overpic}
\usepackage{subcaption}
\usepackage{adjustbox}

\usepackage[letterpaper,top=2cm,bottom=2cm,left=3cm,right=3cm,marginparwidth=1.75cm]{geometry}

\usepackage{soul}
\usepackage{amsmath}
\usepackage{amssymb}

\usepackage[numbers]{natbib}

\usepackage{graphicx}
\usepackage[colorlinks=true, allcolors=blue]{hyperref}

\begin{document}

\begin{frontmatter}
\title{Deep Gaussian Process-based Cost-Aware Batch Bayesian Optimization for Complex Materials Design Campaigns}
\author{Sk Md Ahnaf Akif Alvi$^{a,e}$} 
\corref{mycorrespondingauthor}
\ead{ahnafalvi@tamu.edu}
\author{Brent Vela$^{a}$}
\author{Vahid Attari$^{a}$}

\author{Jan Janssen$^{d}$}

\author{Danny Perez$^{e}$}
\author{Douglas Allaire$^{b}$}

\author{Raymundo Arróyave$^{a,b,c}$}
    
\address{$^a$Department of Materials Science and Engineering, Texas A\&M University, College Station, TX, USA 77843}

\address{$^b$ J. Mike Walker '66 Department of Mechanical Engineering, Texas A\&M University, College Station, TX, USA 77843}
\address{$^c$ Wm Michael Barnes '64 Department of Industrial and Systems Engineering, Texas A\&M University, College Station, TX, USA 77843}
\address{$^d$ Max-Planck-Institute for Sustainable Materials, D\"usseldorf, Germany, 40237}
\address{$^e$ Theoretical Division T-1, Los Alamos National Laboratory, Los Alamos, NM, USA 87544}

\begin{abstract}

The accelerating pace and expanding scope of materials discovery demand optimization frameworks that efficiently navigate vast, nonlinear design spaces while judiciously allocating limited evaluation resources. We present a cost-aware, batch Bayesian optimization scheme powered by deep Gaussian process (DGP) surrogates and a heterotopic querying strategy. Our DGP surrogate, formed by stacking GP layers, models complex hierarchical relationships among high-dimensional compositional features and captures correlations across multiple target properties, propagating uncertainty through successive layers. We integrate evaluation cost into an upper-confidence-bound acquisition extension, which, together with heterotopic querying, proposes small batches of candidates in parallel, balancing exploration of under-characterized regions with exploitation of high-mean, low-variance predictions across correlated properties. Applied to refractory high-entropy alloys for high-temperature applications, our framework converges to optimal formulations in fewer iterations with cost-aware queries than conventional GP-based BO, highlighting the value of deep, uncertainty-aware, cost-sensitive strategies in materials campaigns.
\end{abstract}


\end{frontmatter}

\section{Introduction}

The discovery of new materials, especially structural alloys, is increasingly dependent on combining experiments, simulations, and data-driven methods in a coordinated and efficient way. The Materials Genome Initiative (MGI) emphasized this need, with the aim of accelerating the development of materials by integrating these different approaches \cite{dePablo2019}. In this paradigm, Bayesian optimization (BO)~\cite{shahriari2015taking, frazier2018tutorial} has emerged as a key strategy for steering experiments and computations toward optimal material candidates in a data-efficient way~\cite{arroyave2022perspective}. BO is particularly well suited for materials design problems where evaluating the objective (e.g., synthesizing or simulating a material) is costly and where data are sparse \cite{snoek2012practical, frazier2018tutorial}. In a BO framework, a probabilistic surrogate model (traditionally a Gaussian process, GP) is used to model the material property of interest based on available data, while an acquisition function (e.g., expected improvement or hypervolume improvement) uses the surrogate’s predictions and uncertainties to suggest the next experiments or computations \cite{snoek2012practical, frazier2018tutorial, Rasmussen2006}. This iterative refine-and-query approach efficiently balances exploration and exploitation, accelerating the identification of promising new materials while minimizing the number of expensive trials \cite{snoek2012practical, frazier2018tutorial}. BO’s benefits have been demonstrated across various materials domains, including the optimization of high-entropy alloys (HEAs) where vast compositional design spaces make exhaustive search infeasible \cite{alvi2025deep, alvi2025accurate}. In such contexts, multi-objective goals (e.g. maximizing mechanical performance while minimizing cost or weight) are common, motivating advanced acquisition strategies such as expected hypervolume improvement (EHVI)~\cite{zhao2018fast} for Pareto optimization. Recent developments in BO have extended EHVI to parallel batch selection (the q-EHVI algorithm) to efficiently explore trade-offs among multiple objectives with batch experiments \cite{Daulton2020, Foumani2023}.

Despite the success of GP-based BO in materials science \cite{arroyave2022perspective,hastings2025accelerated,paramore2025two}, conventional GPs face challenges in modeling the rich, hierarchical structure of materials data. Many material systems exhibit highly nonlinear composition–property relationships, heterogeneity in data sources (experimental vs. computational), and inherent noise or heteroscedasticity (property uncertainties varying across composition) \cite{Damianou2013, alvi2025accurate}. Deep Gaussian Processes (DGPs) have emerged as a powerful alternative surrogate modeling approach to address these challenges. A DGP is a hierarchical, multilayered GP that can capture complex, nonstationary functions by composing multiple GP layers \cite{Damianou2013, alvi2025deep, alvi2025accurate}. Intuitively, each layer of a DGP can learn a latent representation of the inputs, allowing subsequent layers to model higher-level abstractions. This added depth enables DGPs to naturally model phenomena like input warping, varying length scales, or multi-modal behaviors that a single-layer GP might struggle with \cite{Damianou2013, alvi2025accurate}. Importantly, DGPs retain the principled uncertainty quantification of GPs, making them uncertainty-aware surrogates, which is a crucial trait for BO where decisions are made under uncertainty. Recent studies in computational materials science have shown that hierarchical DGP models significantly outperform conventional GPs on complex materials data sets, especially when the data are noisy, sparse, or non-uniform \cite{alvi2025accurate, alvi2025deep}. For example, Alvi \textit{ et al.} demonstrated that a DGP augmented with a learned neural network prior achieved more accurate and uncertainty-aware predictions of multiple HEA properties than single-task GPs or deterministic machine learning models \cite{alvi2025accurate}. The DGP’s ability to effectively capture property-property correlations and heteroscedastic behaviors led to robust and data-efficient performance in that study \cite{alvi2025accurate}. These findings underscore that DGP surrogates can provide a hierarchical and uncertainty-aware modeling of materials systems, which is advantageous for guiding optimization.

A key feature of many modern materials optimization campaigns is the availability of multiple data sources or tasks that describe the performance of a material. For example, one may have a mix of high‑fidelity but costly experimental measurements and low‑cost proxy experiments that approximate the target properties at reduced fidelity. An example of this would be the yield strength and hardness \cite{alvi2025accurate, Khatamsaz2023, vela2023data}. Bayesian optimization in such settings calls for heterogeneous or multi-task query strategies that can leverage all sources efficiently. Traditional BO implementations often treat each objective or task independently, using separate GPs for each property. However, this independence assumption neglects valuable information: material properties are frequently correlated (again, yield strength and hardness, density and melting temperature, etc.). Indeed, recent work shows that treating multi-objective materials optimization as a set of independent single-objective problems is suboptimal \cite{Khatamsaz2023}. In contrast, multi-output GPs such as co-kriging or Multi-Task GPs (MTGPs) and deep multi-task models can exploit property-property correlations to improve predictions across all tasks \cite{Bonilla2008}. DGPs naturally extend this concept by learning a shared latent space for multiple outputs, enabling information transfer between tasks even when their data are heterotopic---i.e., not all tasks are observed for all samples. In materials research, heterotopic data situations are common, e.g., hardness experiments are performed for all synthesized alloys, but tensile tests are only performed on the hardest subset of alloys \cite{acemi2024multi}. A DGP-based surrogate can seamlessly handle such incomplete, heterotopic datasets by leveraging the correlations among tasks to fill in informational gaps \cite{alvi2025accurate}. This capability means that DGPs make maximal use of both high- and low-fidelity data to improve model fidelity. Furthermore, this can be extended to exploit correlations among hybrid data, i.e., sparse experimental data and abundant but approximate computational data.

Building on these advances, we introduce a cost-aware batch Bayesian optimization framework that uses deep Gaussian process surrogates to speed up materials discovery. The method targets settings with heterotopic, multi‑fidelity data, where each information source carries a different query cost. Specifically, we simultaneously optimize the performance of a material based on both expensive and cheaper queries, on a limited budget. Our approach employs a multi-output DGP as the surrogate model, providing a hierarchical, uncertainty-aware prediction of all relevant material properties (objectives) across the compositional design space. This surrogate is tightly integrated with a cost-aware acquisition strategy. We use a multi-objective batch acquisition function, namely the q-Expected Hypervolume Improvement (qEHVI), to select batches of candidate materials for evaluation in each iteration \cite{Daulton2020}. The qEHVI criterion allows us to rigorously quantify the expected gain in Pareto front hypervolume from evaluating a set of $q$ new points in parallel, thus naturally handling trade-offs between multiple objectives (e.g., maximizing a property while minimizing the cost of querying) and exploiting parallel experimentation or computation.


To account for differing evaluation costs across information sources, we extend the q-EHVI acquisition function to incorporate cost-weighted utility during batch selection. This cost-aware strategy enables the optimizer to favor inexpensive queries—such as CALPHAD or low-fidelity surrogate outputs—for broad exploration, while reserving costly evaluations, including targeted experiments or high-fidelity computational calculations. For example, our method might first run several inexpensive queries before committing resources to a single expensive query. This strategy leverages differential costs to achieve more information gain per budget spent, making overall optimization more efficient and cost-aware. By dynamically balancing exploratory queries (often cheap, broad surveys via computation) and exploitative queries (targeted, expensive experiments), our framework aims to identify top-performing material candidates with significantly reduced total cost and time. We benchmark our DGP-driven BO approach against the conventional BO. In this baseline, independent single-task GPs model each objective, and only qEHVI acquisition functions guide isotropic query batches without cost differentiation. Consistent with prior observations, we expect the baseline to underperform in complex, multi-objective scenarios where ignoring correlations and cost leads to suboptimal decisions. In contrast, our proposed DGP-BO method, by exploiting shared trends and incorporating cost-awareness, is anticipated to more efficiently navigate the search space.

We deploy this framework in a multi-objective high-entropy alloy (HEA) discovery case study, aiming to efficiently identify compositions that offer improved trade-offs between mechanical and thermal performance. By using a DGP to combine knowledge from expensive and cheap data sources within our batch strategy, and by intelligently allocating the query budget through heterotopic querying, we accelerate materials discovery beyond traditional approaches. We demonstrate that combining hierarchical, uncertainty-aware DGP surrogates with heterotopic query strategies and multi-objective batch acquisitions can substantially outperform classical GP-BO approaches in identifying optimal materials. The framework not only yields more accurate surrogate models for complex material systems but also drives experimental campaigns more economically – a critical step toward realizing the MGI vision of faster, cheaper materials innovation.

The remainder of this paper is organized as follows. \textbf{Section 2} introduces the methodological framework in detail, including the formulation of the cost-aware acquisition strategy. \textbf{Section 3} presents the results of the multi-objective HEA discovery campaign and analyzes the performance of the proposed approach. Finally, \textbf{Section 4} summarizes our key findings and outlines directions for future research.

\section{Methodology}
\subsection{Conventional Gaussian Process Baseline (cGP)}
We model each objective $f(\mathbf{x})$ independently as a Gaussian process (GP), placing a GP prior on
\begin{equation}
    f(\mathbf{x}) \;\sim\; \mathcal{GP}\bigl(m(\mathbf{x}),\,k(\mathbf{x},\mathbf{x}')\bigr),
\end{equation}
with mean function \(m(\mathbf{x})= 0\) and a Matérn-\(\tfrac52\) covariance
\begin{equation}
    k(\mathbf{x},\mathbf{x}') \;=\; \sigma_f^2 
    \!\Bigl(1 + \frac{\sqrt{5}\,\|\mathbf{x}-\mathbf{x}'\|}{\ell} + \frac{5\,\|\mathbf{x}-\mathbf{x}'\|^2}{3\,\ell^2}\Bigr)
    \exp\!\Bigl(-\frac{\sqrt{5}\,\|\mathbf{x}-\mathbf{x}'\|}{\ell}\Bigr),
\end{equation}
where \(\sigma_f^2\) and \(\ell\) are the signal variance and length-scale, respectively \cite{Rasmussen2006}.  
Given data \(\mathcal{D}=\{\mathbf{X}_N,\mathbf{y}_N\}\), the joint distribution of training outputs \(\mathbf{y}_N\) and a new test output \(f(\mathbf{x}^*)\) is
\[
\begin{bmatrix}
    \mathbf{y}_N \\ f(\mathbf{x}^*)
\end{bmatrix}
\sim
\mathcal{N}\!\Biggl(
    \mathbf{0},
    \begin{bmatrix}
        \mathbf{K}_{NN} + \sigma_n^2\mathbf{I} & \mathbf{k}_* \\
        \mathbf{k}_*^\top & k_{**}
    \end{bmatrix}
\Biggr),
\]
with \(\mathbf{K}_{NN}=k(\mathbf{X}_N,\mathbf{X}_N)\), \(\mathbf{k}_*=k(\mathbf{X}_N,\mathbf{x}^*)\), and \(k_{**}=k(\mathbf{x}^*,\mathbf{x}^*)\).  Conditioning (i.e., training the GP) yields the predictive mean and variance \cite{Rasmussen2006}:
\begin{align}
    \mu(\mathbf{x}^*) 
    &= \mathbf{k}_*^\top\bigl(\mathbf{K}_{NN} + \sigma_n^2\mathbf{I}\bigr)^{-1}\mathbf{y}_N, 
    \\[1ex]
    \sigma^2(\mathbf{x}^*) 
    &= k_{**} - \mathbf{k}_*^\top\bigl(\mathbf{K}_{NN} + \sigma_n^2\mathbf{I}\bigr)^{-1}\mathbf{k}_*.
\end{align}
Hyperparameters \(\{\sigma_f^2,\ell,\sigma_n^2\}\) are learned by maximizing the log marginal likelihood:
\begin{equation}
    \log p(\mathbf{y}_N\mid\mathbf{X}_N)
    = -\tfrac12\,\mathbf{y}_N^\top\bigl(\mathbf{K}_{NN} + \sigma_n^2\mathbf{I}\bigr)^{-1}\mathbf{y}_N
      -\tfrac12\log\bigl|\mathbf{K}_{NN} + \sigma_n^2\mathbf{I}\bigr|
      -\tfrac{N}{2}\log 2\pi.
\end{equation}
This framework provides closed‐form uncertainty quantification and scales cubically in \(N\), motivating our investigation of more scalable, expressive surrogates.

\subsection{Deep Gaussian Process (DGP)}
Deep Gaussian processes (DGPs) extend GPs by composing multiple latent GP layers, capturing more complex, non‐stationary patterns and hierarchical features \cite{Damianou2013,salimbeni2017doubly}.  For a two‐layer DGP, we define
\begin{align}
    \mathbf{h}^{(1)} &= f^{(1)}(\mathbf{x}) + \epsilon^{(1)}, 
    &\epsilon^{(1)} &\sim \mathcal{N}\bigl(\mathbf{0},\sigma_1^2\mathbf{I}\bigr),
    \\[0.5ex]
    \mathbf{y} &= f^{(2)}\bigl(\mathbf{h}^{(1)}\bigr) + \epsilon^{(2)},
    &\epsilon^{(2)} &\sim \mathcal{N}\bigl(\mathbf{0},\sigma_2^2\mathbf{I}\bigr),
\end{align}
where each \(f^{(\ell)}\) is a GP with its own kernel \(k^{(\ell)}\), inducing inputs \(\mathbf{Z}^{(\ell)}\), and variational parameters \(\mathbf{m}^{(\ell)},\mathbf{S}^{(\ell)}\).  The exact marginalization over \(\mathbf{h}^{(1)}\) is intractable; therefore, we maximize the doubly stochastic evidence lower bound (ELBO):
\begin{equation}
    \mathcal{L}
    = \sum_{n=1}^N \mathbb{E}_{q(\mathbf{h}_n^{(1)})}\bigl[\log p(y_n\!\mid\!\mathbf{h}_n^{(1)})\bigr]
      \;-\; \sum_{\ell=1}^2 \mathrm{KL}\bigl[q(\mathbf{u}^{(\ell)}) \,\|\, p(\mathbf{u}^{(\ell)})\bigr],
\end{equation}
where \(q(\mathbf{u}^{(\ell)})=\mathcal{N}(\mathbf{m}^{(\ell)},\mathbf{S}^{(\ell)})\) approximates the prior over inducing outputs \cite{salimbeni2017doubly}.  We implement 32 inducing points per layer and 10 latent GPs in the first layer, optimizing \(\mathcal{L}\) via Adam (lr=0.01) with early stopping after 50 epochs.  This hierarchical construction allows the model to learn input‐dependent warping of the feature space, effectively capturing non‐Gaussian predictive distributions and multi‐scale correlations.

\subsection{Acquisition Functions}
We employ a combination of cost-aware single-objective and multi-objective acquisition functions to balance exploration, exploitation, and evaluation cost across heterogeneous data sources as described below.

\subsubsection{Upper Confidence Bound (UCB)}
For cost‐aware single‐objective sampling, we use UCB, defined below:
\[
\mathrm{UCB}(x) = \mu(x) + \beta\,\sigma(x),
\]
with \(\beta\) tuning the exploration–exploitation balance \cite{wilson2018maximizing,srinivas2009gaussian}.  We draw candidate tasks proportionally to a softmax over their evaluation costs and optimize UCB on a discrete candidate set via a simulated annealing optimizer.

\subsubsection{q-Expected Hypervolume Improvement (qEHVI)}
To guide multi-objective batch selection, we use the qEHVI acquisition function \cite{Daulton2020,emmerich2008computation}, which measures the expected increase in Pareto front hypervolume when evaluating a batch \(\{\mathbf{x}_1,\dots,\mathbf{x}_q\}\) in parallel.  
Let \(P \subset \mathbb{R}^m\) denote the current Pareto set and \(\mathbf{y}_\mathrm{ref}\) a reference point. The hypervolume of \(P\) is given by

\[
\mathrm{HV}(P) = \int_{\mathbf{y}\in\mathbb{R}^m} \mathbb{I}\bigl(P \preceq \mathbf{y} \preceq \mathbf{y}_\mathrm{ref}\bigr)\,d\mathbf{y},
\]
and the improvement from adding \(\mathbf{y}\) is \(I(\mathbf{y},P)=\mathrm{HV}(P\cup\{\mathbf{y}\})-\mathrm{HV}(P)\).  qEHVI is then
\[
\mathrm{qEHVI}(\{\mathbf{x}_i\}) 
= \mathbb{E}_{\mathbf{Y}\sim\mathcal{N}(\boldsymbol{\mu},\boldsymbol{\Sigma})}\bigl[I(\mathbf{Y},P)\bigr],
\]
where \(\boldsymbol{\mu},\boldsymbol{\Sigma}\) are the joint predictive mean and covariance of the batch \cite{Daulton2020}.  We compute qEHVI by decomposing the dominated region into axis‐aligned hyperrectangles via a fast partitioning algorithm, evaluating the multivariate Gaussian CDF over each cell, and summing contributions.  Optimization over the batch is performed using the same simulated annealing routine as for UCB, with batch size \(q=5\).

\subsubsection{Heterotopic Querying Strategy}
We alternate UCB‐based heterotopic queries (cheap, single‐objective) and qEHVI‐based multi‐objective batches every third iteration.  At heterotopic steps, we select two moderate‐uncertainty points (\(\beta=0.25\)) and one high‐uncertainty point (\(\beta=1.0\)) per cost‐weighted task, leveraging inexpensive evaluations to reduce uncertainty before invoking costly multi‐objective evaluations \cite{Foumani2023}. For more information on qEHVI please see Ref. \cite{daulton2020differentiable}.


\subsection{Pareto Front Convexity Ratio}
Multi-objective optimization problems can be characterized by the geometric properties of their Pareto fronts. To quantify the convexity of a Pareto front, we introduce the \emph{convexity ratio} as a metric that captures the proportion of Pareto optimal points that lie on the convex hull of the entire Pareto set. The convexity ratio is formally defined as:

\begin{equation}
\text{Convexity Ratio} = \frac{|\mathcal{P}_{\text{hull}}|}{|\mathcal{P}|} = \frac{n_{\text{hull}}}{n_{\text{total}}}
\label{eq:convexity_ratio}
\end{equation}
where, $n_{\text{hull}}$ represents the number of Pareto points on the convex hull,
$n_{\text{total}}$ represents the total number of Pareto optimal points, $\mathcal{P}_{\text{hull}} = \{p \in \mathcal{P} : p \text{ is a vertex of } \text{ConvexHull}(\mathcal{P})\}$ is the subset of Pareto points on convex hull, and
$\mathcal{P}$ is the set of all Pareto optimal points in the objective space. 

The convexity ratio (CR) is a useful scalar metric for characterizing the shape of Pareto sets in multi-objective (multi-dimensional) optimization problems. By construction, its values lie strictly between 0 and 1, with a maximum of CR=1 corresponding to a fully convex Pareto front. This condition holds if and only if all segments of the Pareto set exhibit convex geometry. When the Pareto front contains one or more non-convex regions—such as inflections, concavities, or disconnected segments—the convexity ratio drops below unity, serving as a diagnostic for geometric complexity. Importantly, the metric is monotonic: as the Pareto set becomes increasingly concave or fragmented, the CR decreases accordingly, reflecting the extent to which the front deviates from ideal convexity. CR thus serves as a compact yet informative indicator of global shape of the Pareto front, and can be used to compare optimization runs or guide surrogate-based exploration strategies.

Based on the convexity ratio, multi-objective optimization toy problems in our current work can be classified into distinct categories. Among our toy problems (listed below), the convexity ratios demonstrate varying degrees of front convexity, arranged here in decreasing order of convexity:

\begin{enumerate}
    \item \textbf{Kursawe} ($CR = 0.500$): Moderately convex problem at the boundary between convex and non-convex characteristics
    \item \textbf{ZDT1} ($CR = 0.303$): Moderately non-convex problem with significant concave regions
    \item \textbf{DTLZ2} ($CR = 0.124$): Highly non-convex problem with complex geometric structure
    \item \textbf{Latent-Aware Multi-Output Functions} ($CR = 0.114$): Highly non-convex problem with the most challenging Pareto front geometry
\end{enumerate}

This ordering reveals distinct problem categories: \textbf{Moderately Convex Problems} ($0.4 \leq CR < 0.6$) such as Kursawe, where approximately half of the Pareto points lie on the convex hull; \textbf{Moderately Non-Convex Problems} ($0.2 \leq CR < 0.4$) such as ZDT1, where a significant portion of the Pareto front exhibits non-convex characteristics; and \textbf{Highly Non-Convex Problems} ($CR < 0.2$) such as DTLZ2 and Latent-Aware Multi-Output Functions, where the vast majority of Pareto optimal points are interior to the convex hull, indicating substantial concavity and complex geometric structure.

The convexity ratio can be computed by: (1) identifying all Pareto optimal points $\mathcal{P}$ in the objective space, (2) computing the convex hull of $\mathcal{P}$ using computational geometry algorithms, (3) determining which Pareto points are vertices of the convex hull to form $\mathcal{P}_{\text{hull}}$, and (4) calculating the ratio using Equation~\ref{eq:convexity_ratio}. 

\section{Results and Discussion}
\subsection{Comparison of cGP-BO, hDGP-BO, and hDGP-MTGP-BO on Synthetic Benchmark Functions}
\begin{figure}[H]
  \centering
  \includegraphics[width=\textwidth]{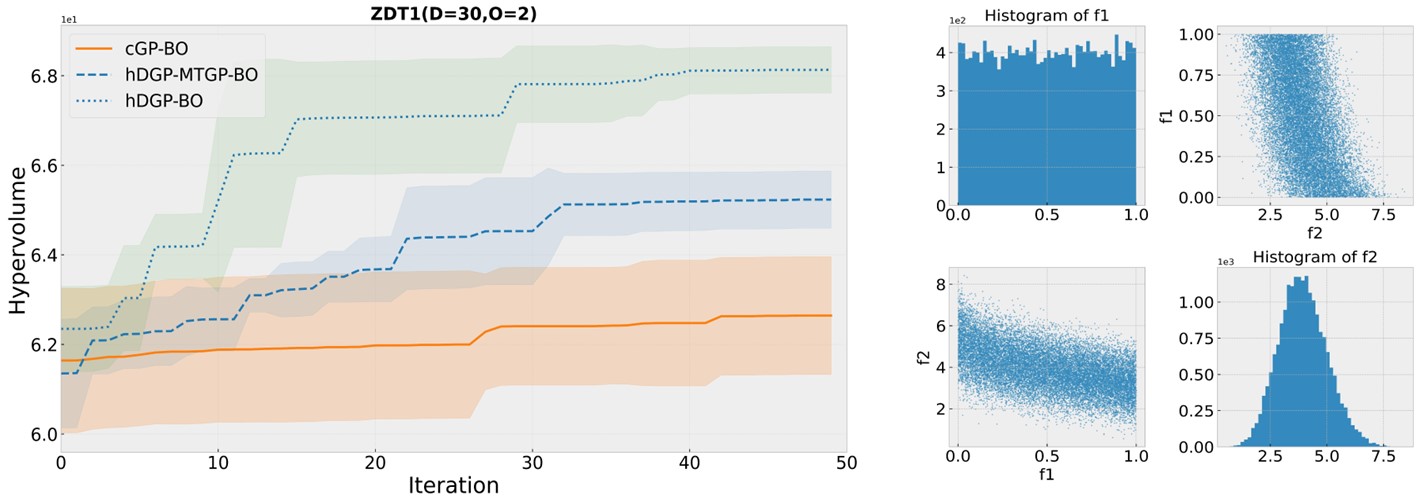}
  \caption{Hypervolume convergence (left) and Pareto front (right) for ZDT1 ($D=30,O=2$).}
  \label{fig:toy4}
\end{figure}

\begin{figure}[H]
  \centering
  \includegraphics[width=\textwidth]{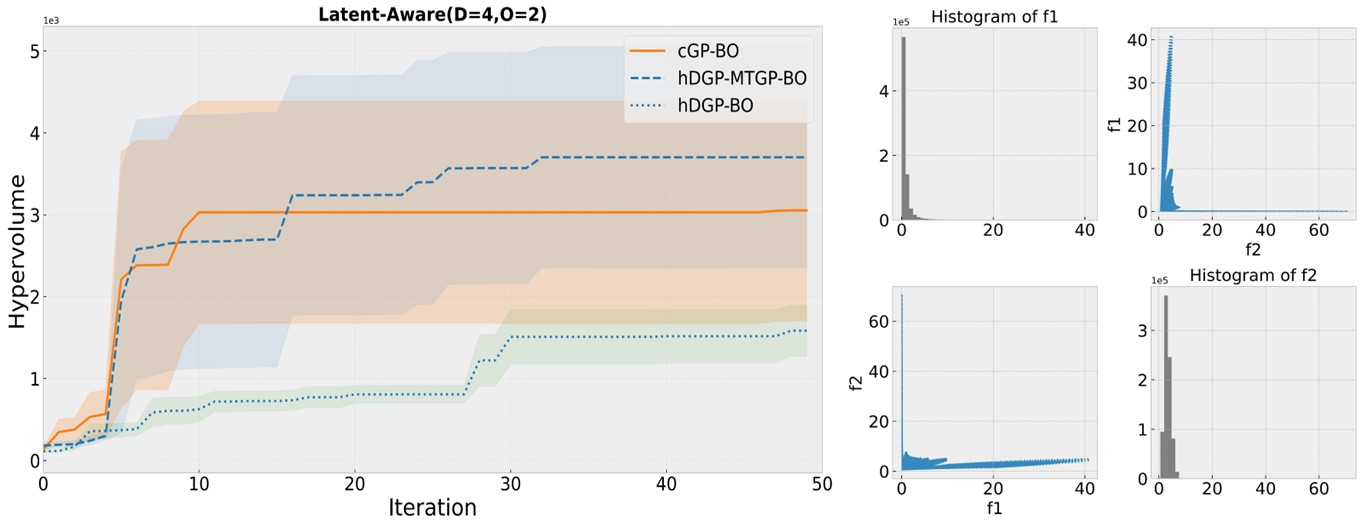}
  \caption{Hypervolume convergence (left) and Pareto‐front scatter plus marginal histograms (right) for the latent‐aware function ($D=4,O=2$).}
  \label{fig:toy1}
\end{figure}

In this subsection, we compare the performance of conventional Gaussian process-based Bayesian optimization (cGP-BO), deep Gaussian process-based Bayesian optimization (hDGP-BO), and our novel hybrid approach, hDGP-MTGP-BO, on four synthetic benchmark functions. Each of these synthetic optimization problems have different pareto front convexity ratios, as defined in Methods. The hDGP-MTGP-BO method combines the mean predictions from hierarchical deep Gaussian processes with variance estimates from multi-task Gaussian processes to address potential limitations in variance predictions arising from double stochastic variational inference \cite{salimbeni2017doubly}. We measure performance by the hypervolume improvement over iterations (left panels in Figures~\ref{fig:toy1}–\ref{fig:toy4}) and characterize the resulting Pareto‐front distributions via scatter plots and marginal histograms (right panels). Cases 1–4 correspond to the functions defined in Equations~\eqref{eq:toy2}–\eqref{eq:toy1}. BO runs for all schemes were averaged over 5 runs, having random starting data.

\subsubsection{Case 1: Kursawe Function ($D=3,O=2$ ; Convexity ratio= 0.500)}\label{sec:toy2}
The Kursawe test problem is defined by\cite{kursawe1991variant}:
\begin{align}\label{eq:toy2}
f_1(\mathbf{x}) &= -10\exp\bigl(-0.2\sqrt{x_1^2 + x_2^2}\bigr)
                   -10\exp\bigl(-0.2\sqrt{x_2^2 + x_3^2}\bigr),\\
f_2(\mathbf{x}) &= |x_1|^{0.8} + 5\sin(x_1^3)
                   + |x_2|^{0.8} + 5\sin(x_2^3)
                   + |x_3|^{0.8} + 5\sin(x_3^3).
\end{align}

\begin{figure}[H]
  \centering
  \includegraphics[width=\textwidth]{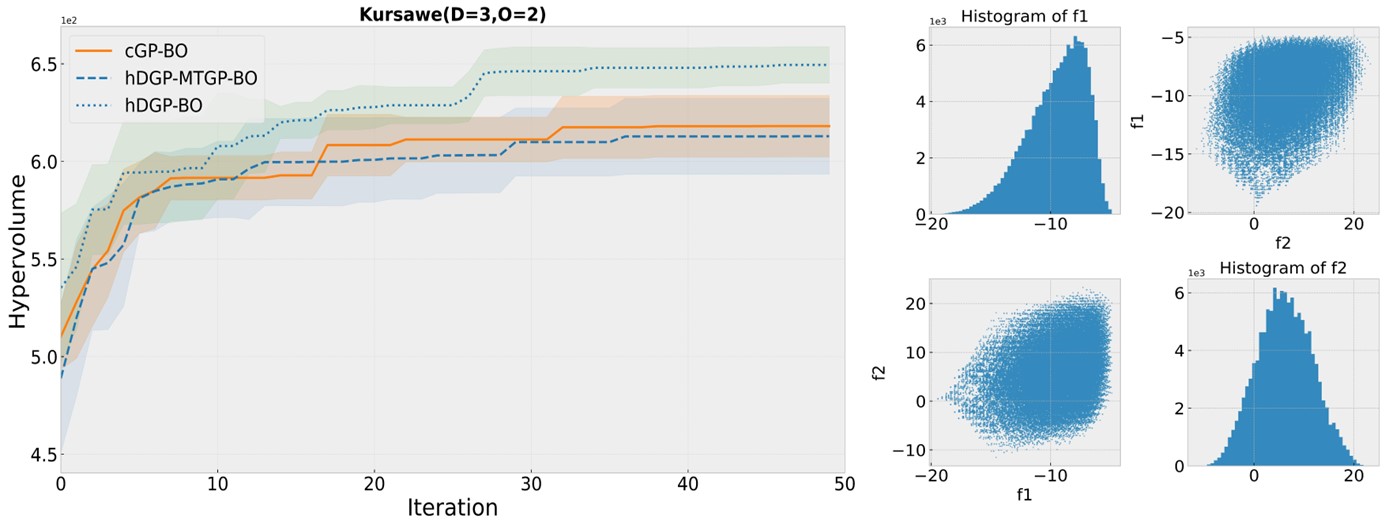}
  \caption{Hypervolume convergence (left) and Pareto‐front analysis (right) for the Kursawe function ($D=3,O=2$).}
  \label{fig:toy2}
\end{figure}

On this nonconvex, low‐dimensional problem, all three methods achieve comparable final hypervolumes, but with distinct convergence characteristics. hDGP-BO shows the most robust behavior throughout optimization. Notably, hDGP-BO achieves the highest final hypervolume ($\approx$6.5$\times$ $10^2$) and demonstrates superior convergence stability compared to hybrid hDGP-MTGP-BO. The hybrid approach provides no modeling advantages over pure DGP implementations. 

\subsubsection{Case 2: ZDT1 Function ($D=30,O=2$; Convexity ratio= 0.303)}\label{sec:toy4}
The high‐dimensional ZDT1 problem is defined by\cite{zitzler2000comparison}:
\begin{align}\label{eq:toy4}
f_1(\mathbf{x}) &= x_1,\\
g(\mathbf{x})   &= 1 + \frac{9}{D-1}\sum_{i=2}^{D} x_i,\\
f_2(\mathbf{x}) &= g(\mathbf{x})\Bigl(1 - \sqrt{\tfrac{f_1(\mathbf{x})}{g(\mathbf{x})}}\Bigr).
\end{align}

In this 30‐dimensional benchmark, hDGP-BO significantly outperforms cGP-BO, achieving a final hypervolume of approximately 6.80$\times$ $10^1$ versus 6.25$\times$ $10^1$. Interestingly, hDGP-MTGP-BO achieves an intermediate performance level ($\approx$6.52$\times$ $10^1$), suggesting a trade‐off between the superior mean prediction capabilities of pure deep GPs and the enhanced stability provided by multi-task variance estimation. The deep surrogate's latent warping enables it to capture the nonlinear decreasing front more effectively in high dimensions, but the hybrid approach shows that combining deep mean predictions with multi-task variance estimates can provide a bit lower performance. 

\subsubsection{Case 3: DTLZ2 Function ($D=7,O=3$; Convexity ratio= 0.124)}\label{sec:toy3}
The three‐objective DTLZ2 benchmark is given by\cite{deb2005scalable}:
\begin{align}\label{eq:toy3}
g(\mathbf{x})     &= \sum_{j=3}^{7} (x_j - 0.5)^2,\\
f_1(\mathbf{x})   &= (1+g)\cos\!\Bigl(\tfrac{\pi}{2}x_1\Bigr)\cos\!\Bigl(\tfrac{\pi}{2}x_2\Bigr),\\
f_2(\mathbf{x})   &= (1+g)\cos\!\Bigl(\tfrac{\pi}{2}x_1\Bigr)\sin\!\Bigl(\tfrac{\pi}{2}x_2\Bigr),\\
f_3(\mathbf{x})   &= (1+g)\sin\!\Bigl(\tfrac{\pi}{2}x_1\Bigr).
\end{align}

\begin{figure}[H]
  \centering
  \includegraphics[width=\textwidth]{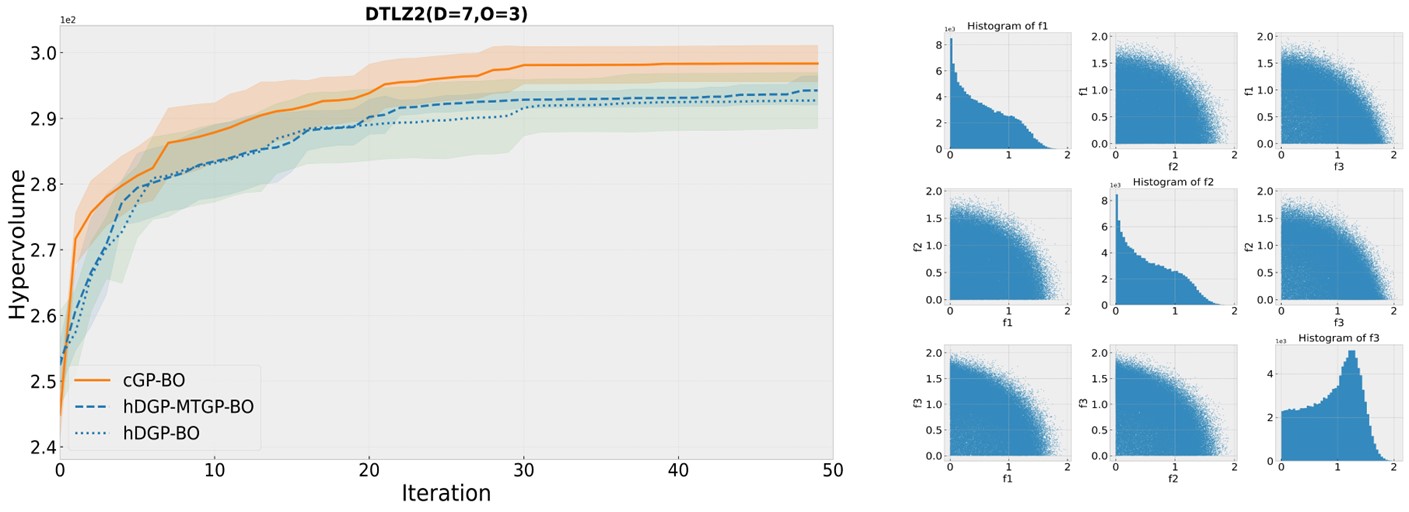}
  \caption{Hypervolume convergence (left) and 3-D Pareto front projections with marginals (right) for DTLZ2 ($D=7,O=3$).}
  \label{fig:toy3}
\end{figure}

Here, cGP-BO slightly outperforms both deep GP variants, reaching the highest hypervolume ($\approx$298) by iteration 50. Between the deep GP approaches, hDGP-MTGP-BO and  hDGP-BO ($\approx$292) shows similar performance, The spherical Pareto front of DTLZ2 is well captured by a single GP layer; the additional complexity of deep architectures incurs  slower convergence for this moderate‐dimensional, smoothly parameterized front. 

\subsubsection{Case 4: Latent‐Aware Multi‐Output Function ($D=4,O=2$; Convexity ratio= 0.114)}\label{sec:toy1}
The first benchmark is a synthetic multi‐output function that combines six base mappings $f_1,\dots,f_6$ into two objectives $y,y'$ \cite{khatamsaz2025microstructure}:
\begin{align}\label{eq:toy1}
f_1(\mathbf{x}) &= x_1^2 + \exp\!\Bigl(-\tfrac{x_2}{x_3}\Bigr),\\
f_2(\mathbf{x}) &= x_1 + x_3,\\
f_3(\mathbf{x}) &= \frac{x_2}{1 + x_3},\\
f_4(\mathbf{x}) &= \log(x_4 + 1)\,x_1,\\
f_5(\mathbf{x}) &= x_2\,\sin(x_4) + \exp(x_1),\\
f_6(\mathbf{x}) &= \sin(x_3) + \cos(x_4),\\
y(\mathbf{x})      &= \frac{1}{10}\Bigl(f_1 f_2 + \tfrac{f_2}{f_3} + f_5 f_4 + f_6\Bigr),\\
y'(\mathbf{x}) &= f_3\,f_2^2 + \frac{f_4}{f_1} + f_5\,f_6.
\end{align}

Early in optimization (iterations 1–10), cGP-BO quickly captures coarse trends, yielding rapid hypervolume gains, but plateaus near 3.0$\times$ $10^3$. In contrast, hDGP-BO, by modeling hierarchical nonlinearities and heteroscedasticity, attains higher final hypervolume ($\approx$1.7$\times$ $10^3$) but exhibits erratic convergence behavior. Most remarkably, hDGP-MTGP-BO achieves the highest final performance ($\approx$3.7$\times$ $10^3$), demonstrating that the hybrid variance estimation approach successfully stabilizes the deep GP benefits while maintaining superior modeling capacity. The improved variance quantification enables more reliable acquisition function evaluations, leading to more consistent exploration of the Pareto front. The skewed marginals in the histograms reflect heavy‐tailed objective distributions that benefit from the deeper latent transformations, with hDGP-MTGP-BO effectively balancing the mean prediction accuracy of deep GPs with robust uncertainty quantification from multi-task learning.

\bigskip
\noindent
In summary, the comparative analysis reveals that both convexity ratio and problem dimensionality critically influence the relative performance of the three approaches. Deep GP surrogates demonstrate clear advantages in two distinct scenarios: high-dimensional problems regardless of convexity (Case 2: ZDT1 with D=30, CR=0.303) and highly non-convex problems with complex function structure (Case 4: Latent-Aware with CR=0.114). Conversely, conventional GPs remain competitive or superior on moderate-dimensional problems with simple geometric structure, even when highly non-convex (Case 3: DTLZ2 with D=7, CR=0.124), due to their lower model complexity and faster convergence. For moderately convex problems (Case 1: Kursawe with CR=0.500), all methods achieve comparable performance regardless of dimensionality. Although the  hybrid HDGP-MTGP-BO is competitive, HDGP-BO perform better overall. This proves that, the uncertainty quantification with variational inference is good enough  for the current problems at hand. \cite{hebbal2019bayesian,kandasamy2015high}.

\subsection{Comparison of cGP-BO and hDGP-BO on in-silico RHEA Design Campaign; convexity ratio =1.00}

Beyond toy benchmarks, we validated hDGP-BO on realistic, high-dimensional design tasks. Here we benchmark hDGP-BO against a conventional GP-BO in batch setting (cGP-BBO) and sequential mode(cGP-SBO) in an in-silico campaign to discover refractory high-entropy alloys (RHEAs) for high-temperature service. In this campaign we explore the Mo–Nb–Cr–Ta–W–Hf alloy space, sampled on a $\Delta c = 5~\text{at.\%}$ grid considering binary to senary alloys, to identify candidate alloys. The optimizer aims to: maximize solidus temperature (thermal stability), yield strength (mechanical performance), thermal conductivity (hot-spot mitigation), and the density (thermal and mechanical performance proxy); also maximizing BCC-content to enforce single-phase BCC stability to avoid deleterious phases. Although no specific component is targeted, the study is motivated by high-temperature applications and is intended to stress-test the algorithm on a complex, multi-objective landscape.

\begin{table}[t]
\centering
\caption{Optimization objectives and constraints for the in-silico RHEA design campaign.}
\begin{tabular}{llp{7.2cm}}
\toprule
\textbf{Property} & \textbf{Target} & \textbf{Rationale} \\
\midrule
Solidus temperature ($T_\text{sol}$) & Maximize & Ensure reliable operation at very high temperatures. \\
Phase stability: single-phase BCC & Maximize & Avoid deleterious secondary phases; target printable, stable solid solutions with BCC phases. \\
Yield strength ($\sigma_y$) & Maximize & Maintain load-bearing capability at service conditions. \\
Thermal conductivity ($\kappa$) & Maximize & Mitigate hot spots and improve thermal management. \\
Density ($\rho$) & Maximize & Dense refractories typically exhibit higher strength, better creep resistance, and higher thermal conductivity at elevated temperatures, crucial for structural applications. \\
Pugh ratio ($B/G$) & Auxiliary Task & Proxy for intrinsic ductility (higher $B/G$ correlates with ductile behavior). \\
Hardness & Auxiliary Task & Screen for high resistance to plastic deformation and wear. \\
\bottomrule
\end{tabular}
\label{tab:objectives}
\end{table}

\begin{figure}[H]
  \centering
  \includegraphics[width=\textwidth]{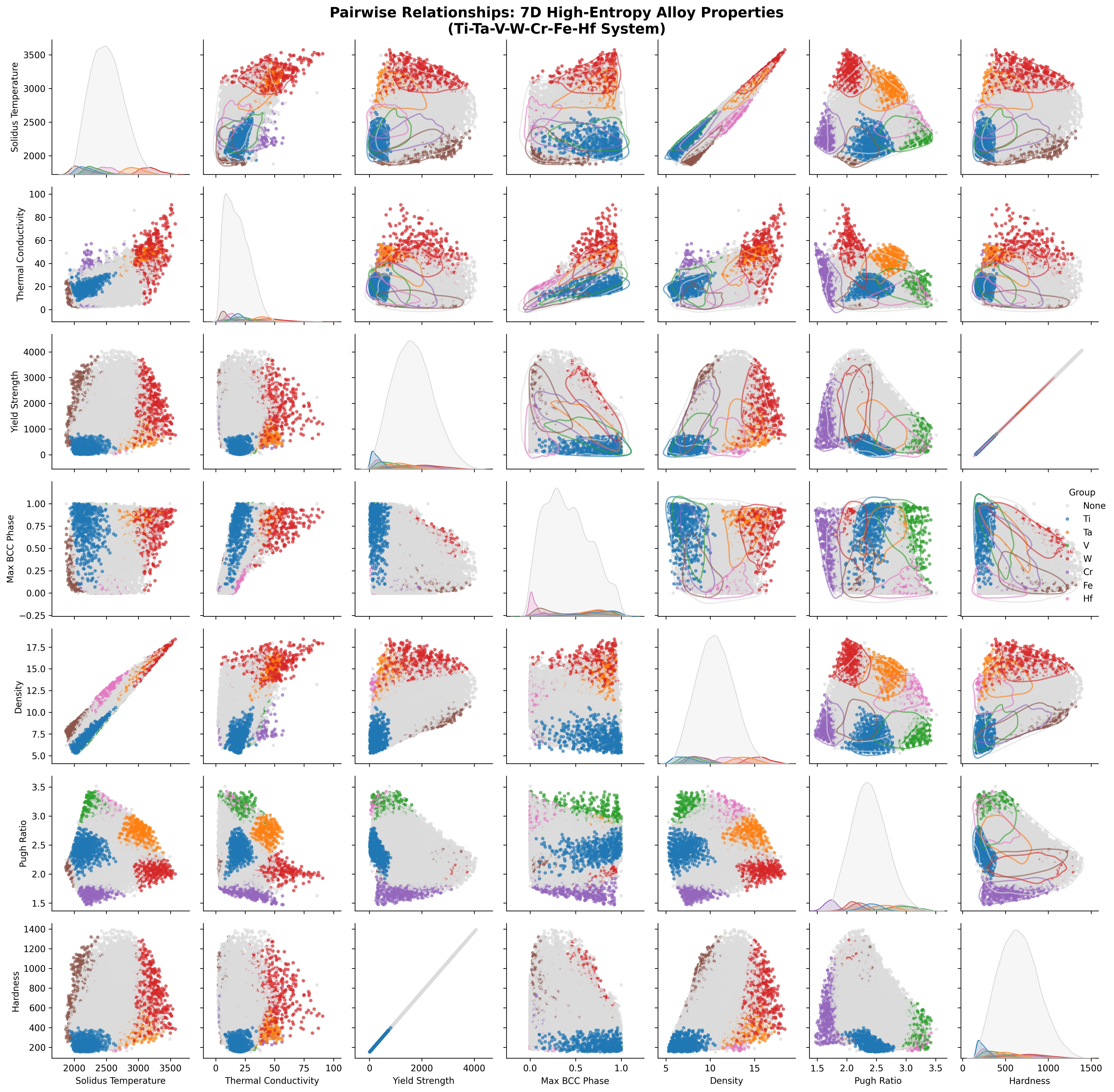}
  \caption{Pairwise relationships among the seven alloy properties.  Diagonal panels show marginal densities; off‐diagonals show scatterplots with group‐wise density contours.}
  \label{fig:pairwise}
\end{figure}

\begin{figure}
  \centering
  \includegraphics[width=0.7\textwidth]{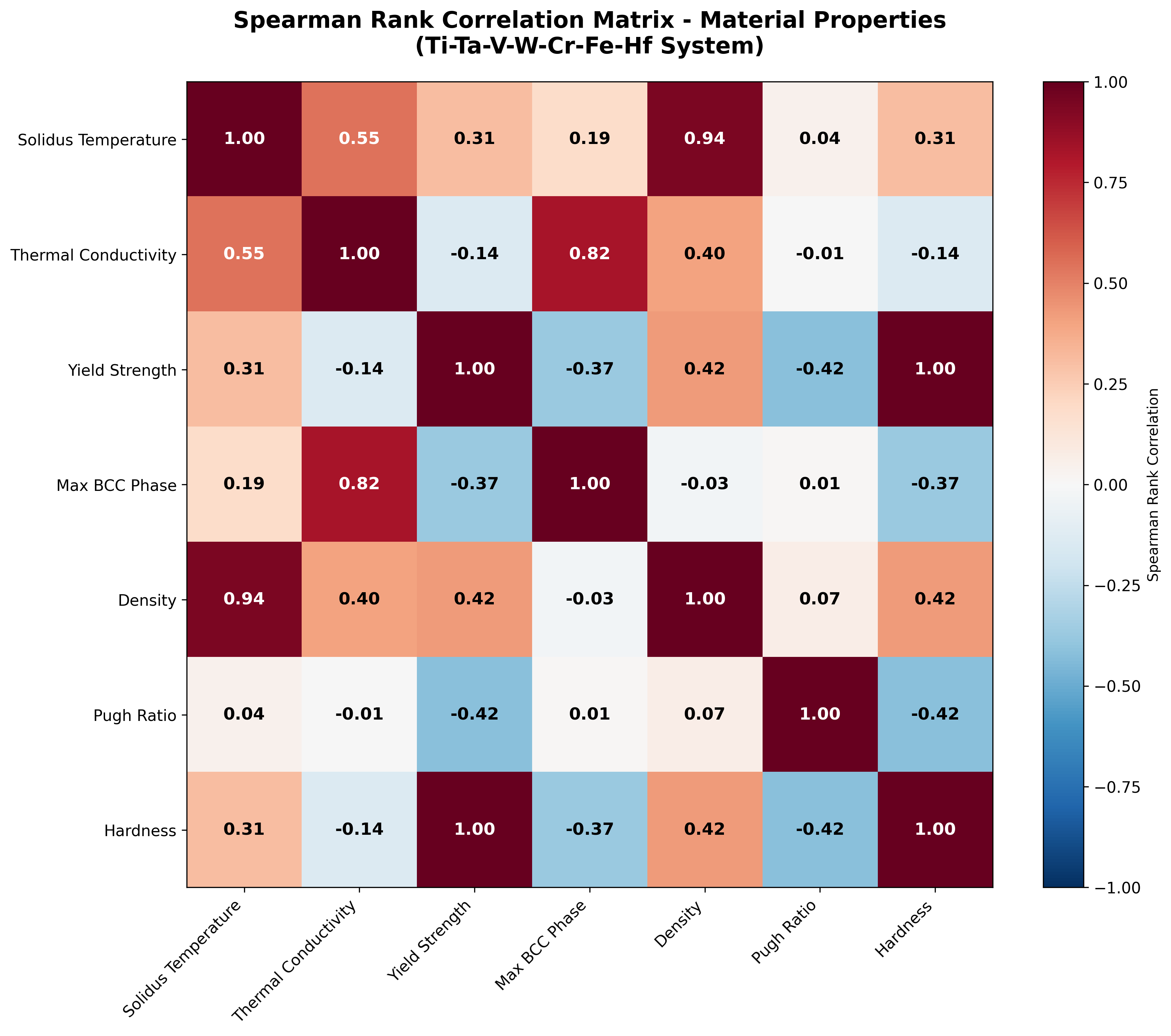}
  \caption{Spearman rank correlation coefficients between properties.  Bold annotations indicate coefficient magnitude.}
  \label{fig:spearman}
\end{figure}

Figure~\ref{fig:pairwise} presents the pairwise scatterplot matrix with marginal densities and group‐wise kernel density contours, revealing both linear and non‐linear dependencies among the seven alloy properties.  In particular, density and solidus temperature exhibit a tight, near‐linear relationship (Spearman \(\rho=0.94\)), reflecting the well‐known influence of atomic packing efficiency on melting behavior in refractory high‐entropy alloys \cite{miracle2017critical}.  Thermal conductivity and maximum BCC phase fraction also correlate strongly (\(\rho=0.82\)), consistent with reports that extended BCC networks facilitate phonon transport in these multicomponent systems \cite{mora2024chemical}.  Conversely, yield strength and Pugh ratio show a broad, scattered relationship, underscoring the complex interplay between elastic moduli and plastic deformation mechanisms in HEAs.  Hardness is highly correlated with yield strength in both Figure~\ref{fig:pairwise} and Figure~\ref{fig:spearman}, because both of these were calculated using Curtin-Maresca model \cite{maresca2020mechanistic} with same theoretical framework.  

The Spearman rank correlation heatmap in Figure~\ref{fig:spearman} quantifies these trends more precisely.  Density–solidus temperature (0.94) and thermal conductivity–BCC phase (0.82) stand out as the strongest positive correlations, confirming the mechanistic coupling of phase stability and transport properties in multinary alloys \cite{zhou2022thermodynamics}.  A negative correlation between Pugh ratio and yield strength (–0.42) highlights the trade‐off between ductility (higher G/B) and strength in BCC‐based systems \cite{pugh1954xcii}.  Other moderate correlations, such as density–yield strength (0.42) and density–hardness (0.42), reflect the contribution of mass density to mechanical reinforcement via solid‐solution strengthening \cite{singh2024alloying}.

\begin{figure}[htbp]
  \centering
  \subfloat[Yield Strength]{%
    \includegraphics[width=0.32\textwidth]{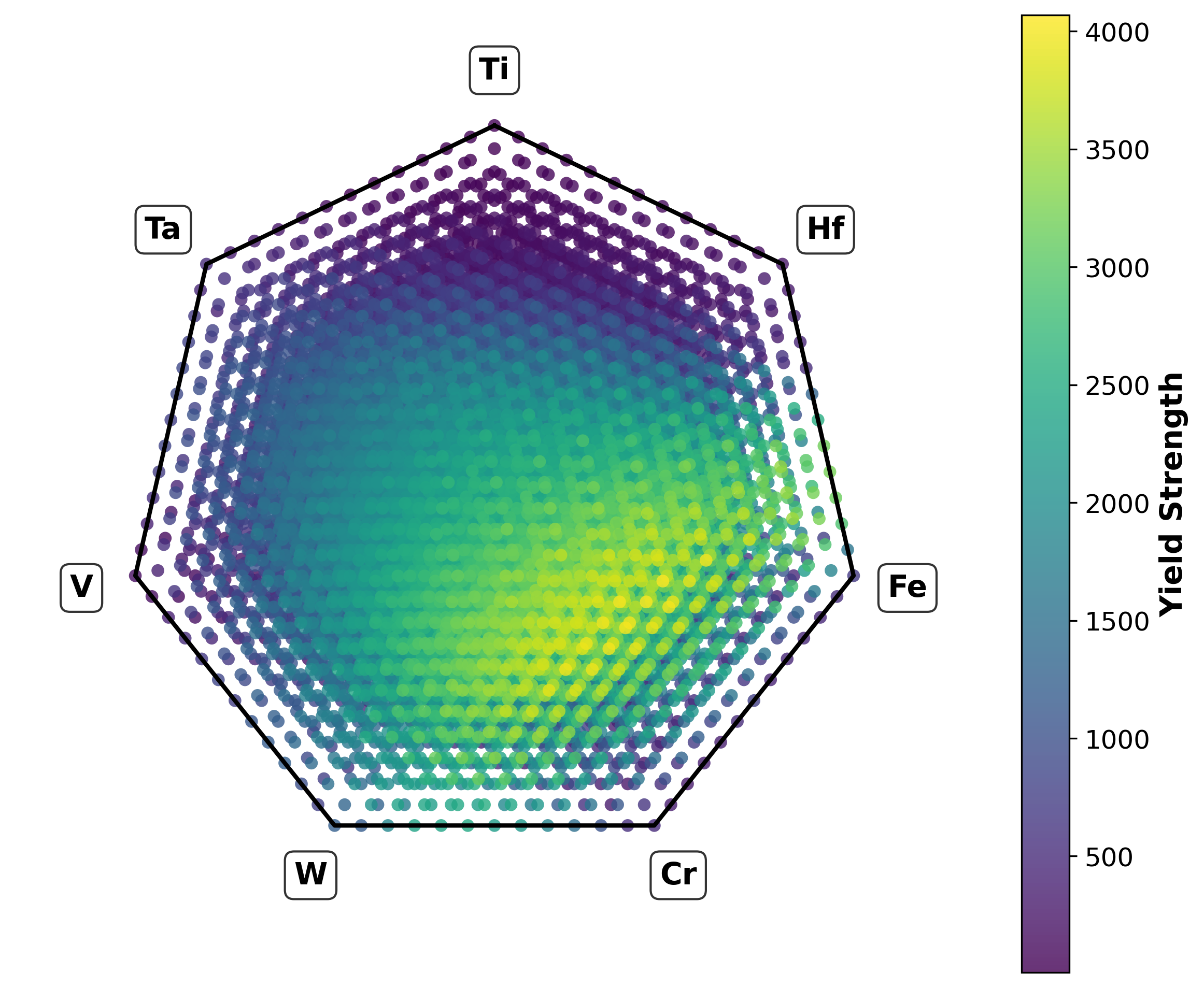}%
  }\hfill
  \subfloat[Thermal Expansion]{%
    \includegraphics[width=0.32\textwidth]{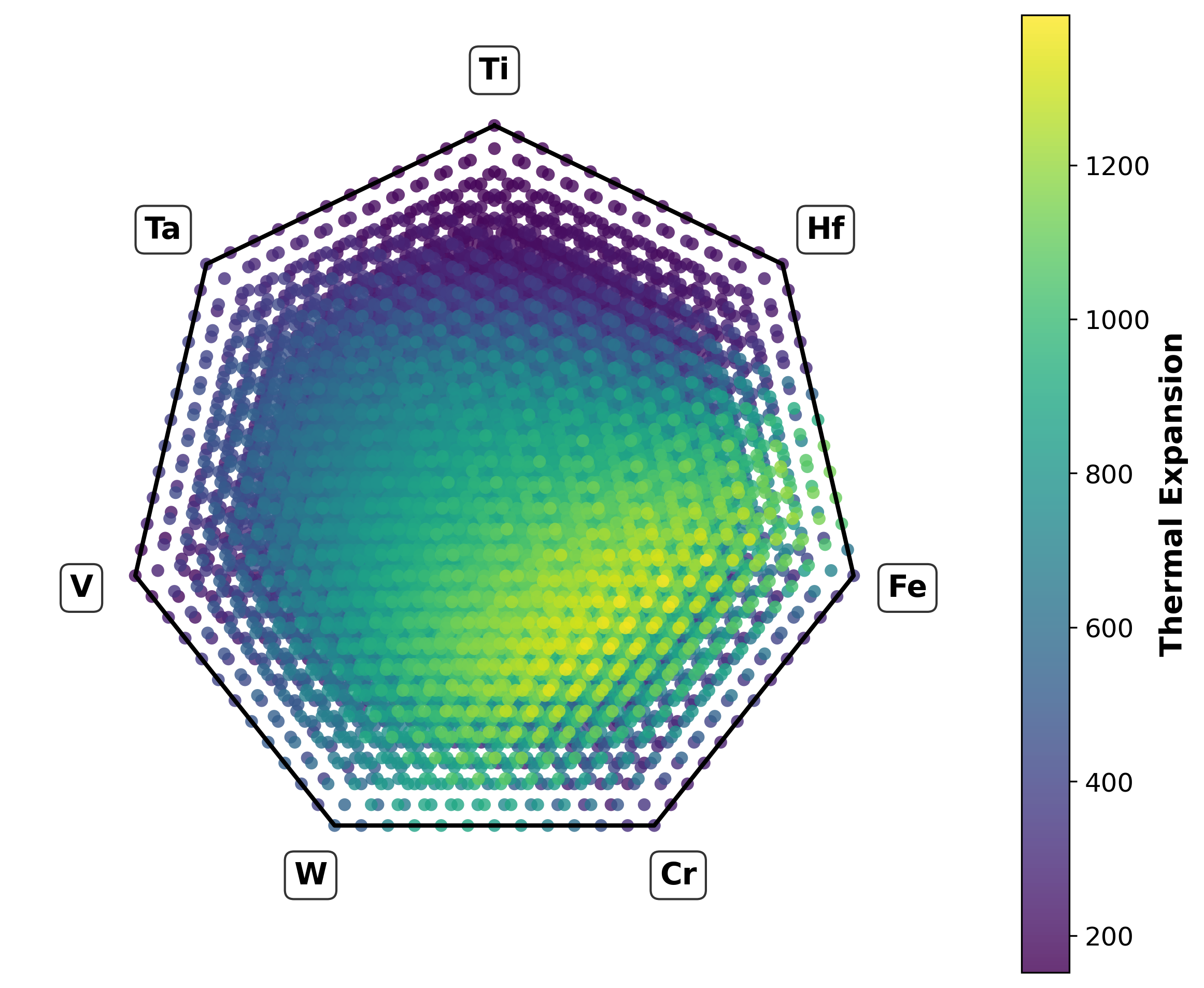}%
  }\hfill
  \subfloat[Thermal Conductivity]{%
    \includegraphics[width=0.32\textwidth]{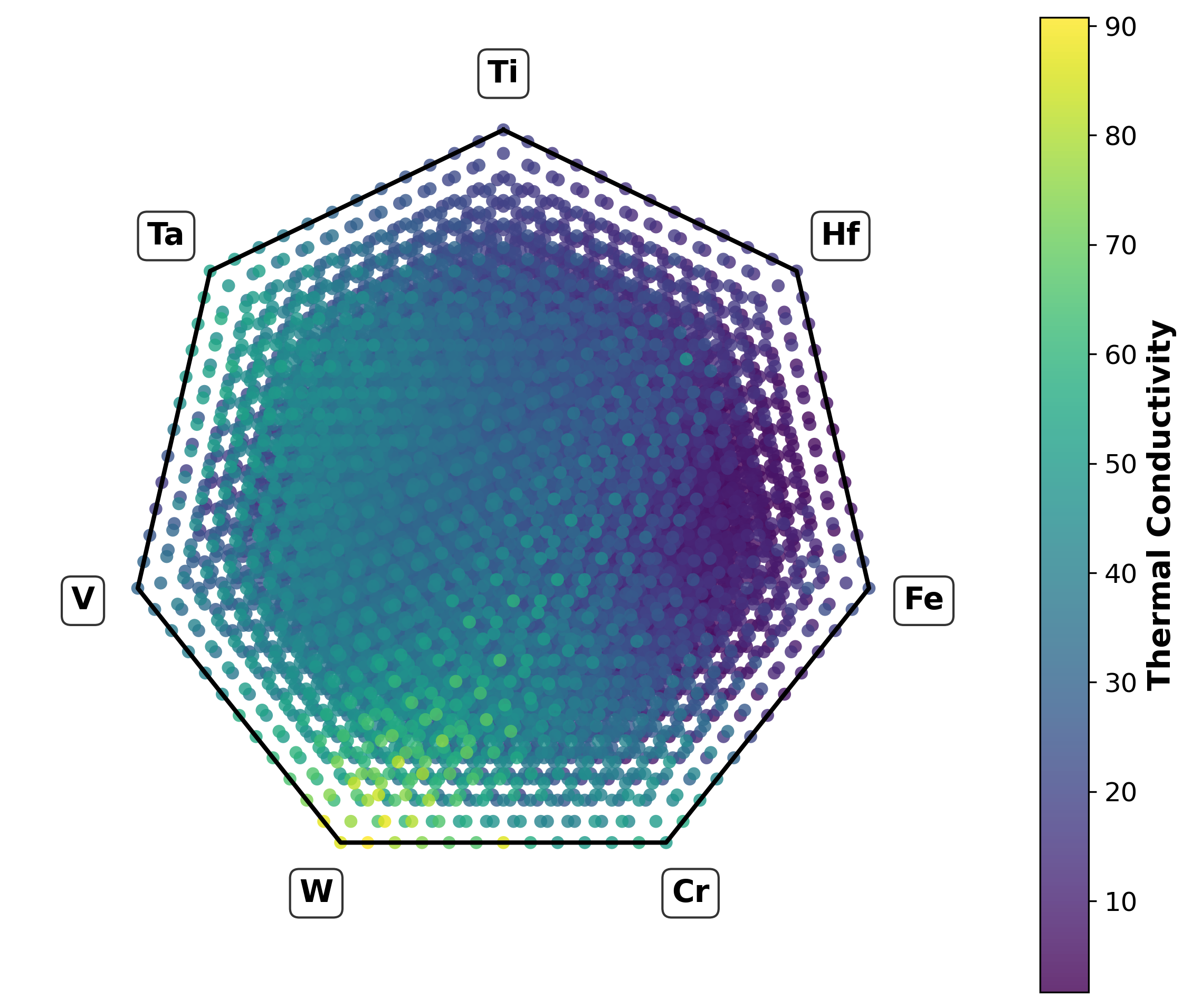}%
  }\\[1ex]
  \subfloat[Solidus Temperature]{%
    \includegraphics[width=0.32\textwidth]{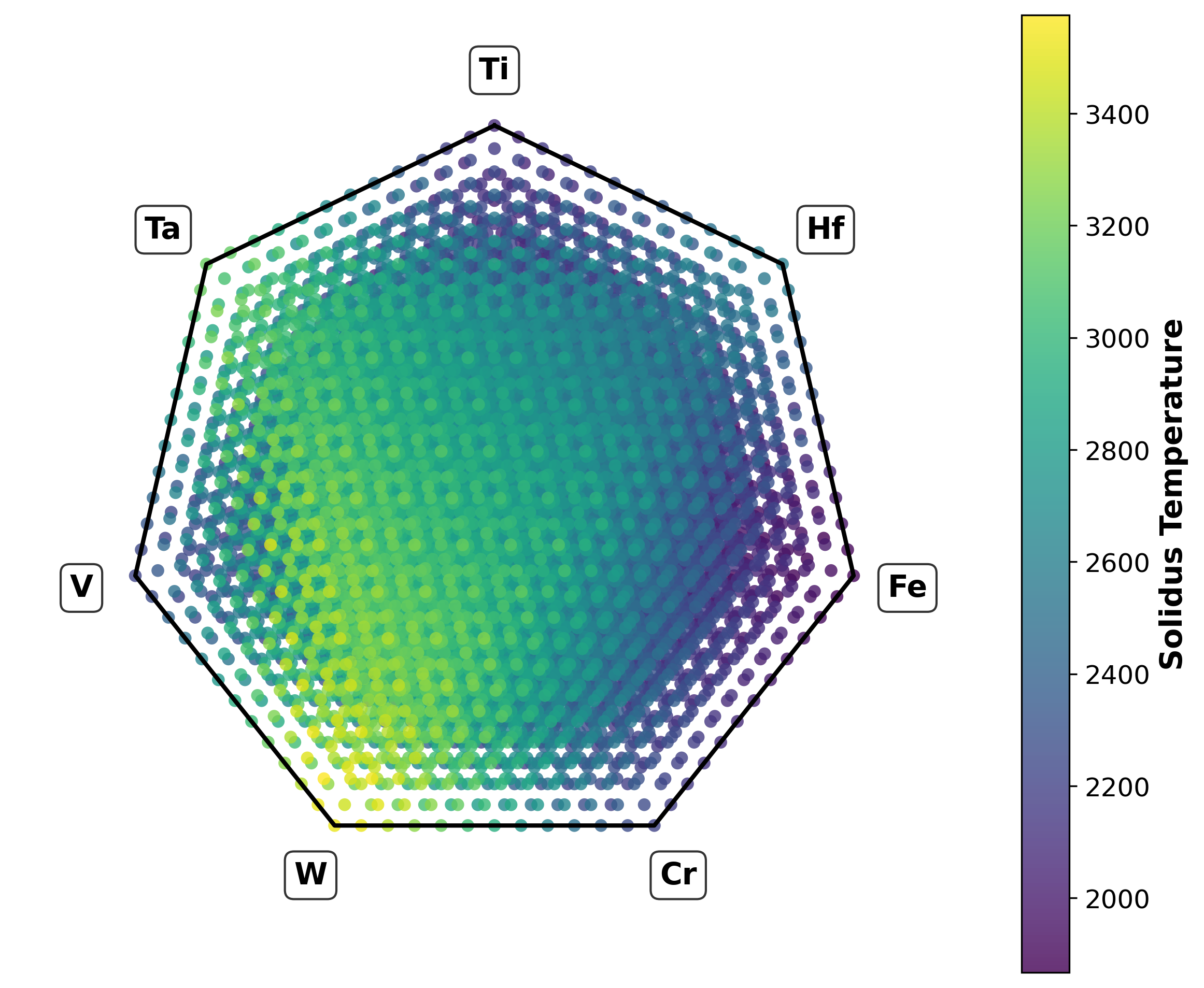}%
  }\hfill
  \subfloat[Pugh Ratio]{%
    \includegraphics[width=0.32\textwidth]{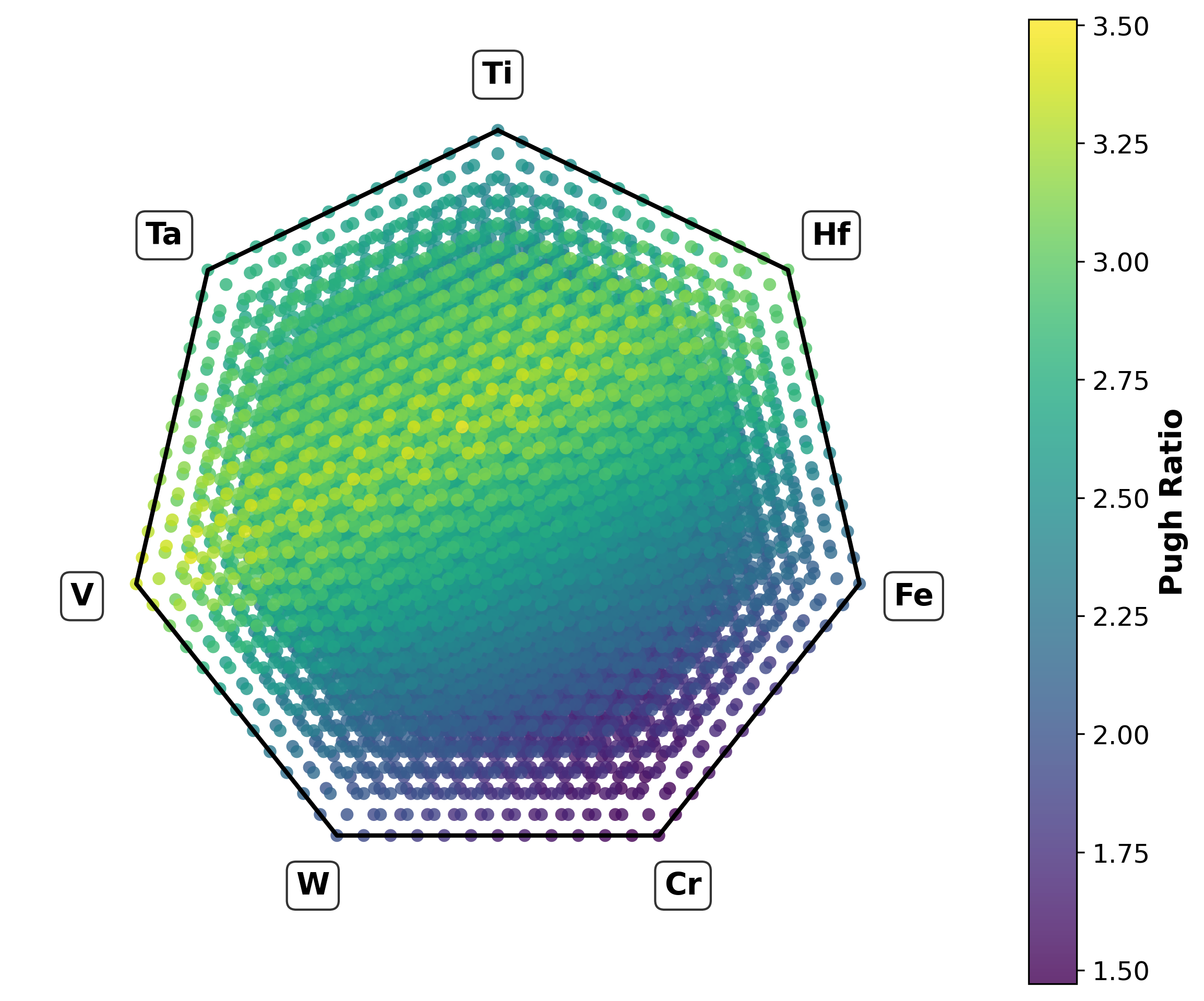}%
  }\hfill
  \subfloat[Max BCC Phase]{%
    \includegraphics[width=0.32\textwidth]{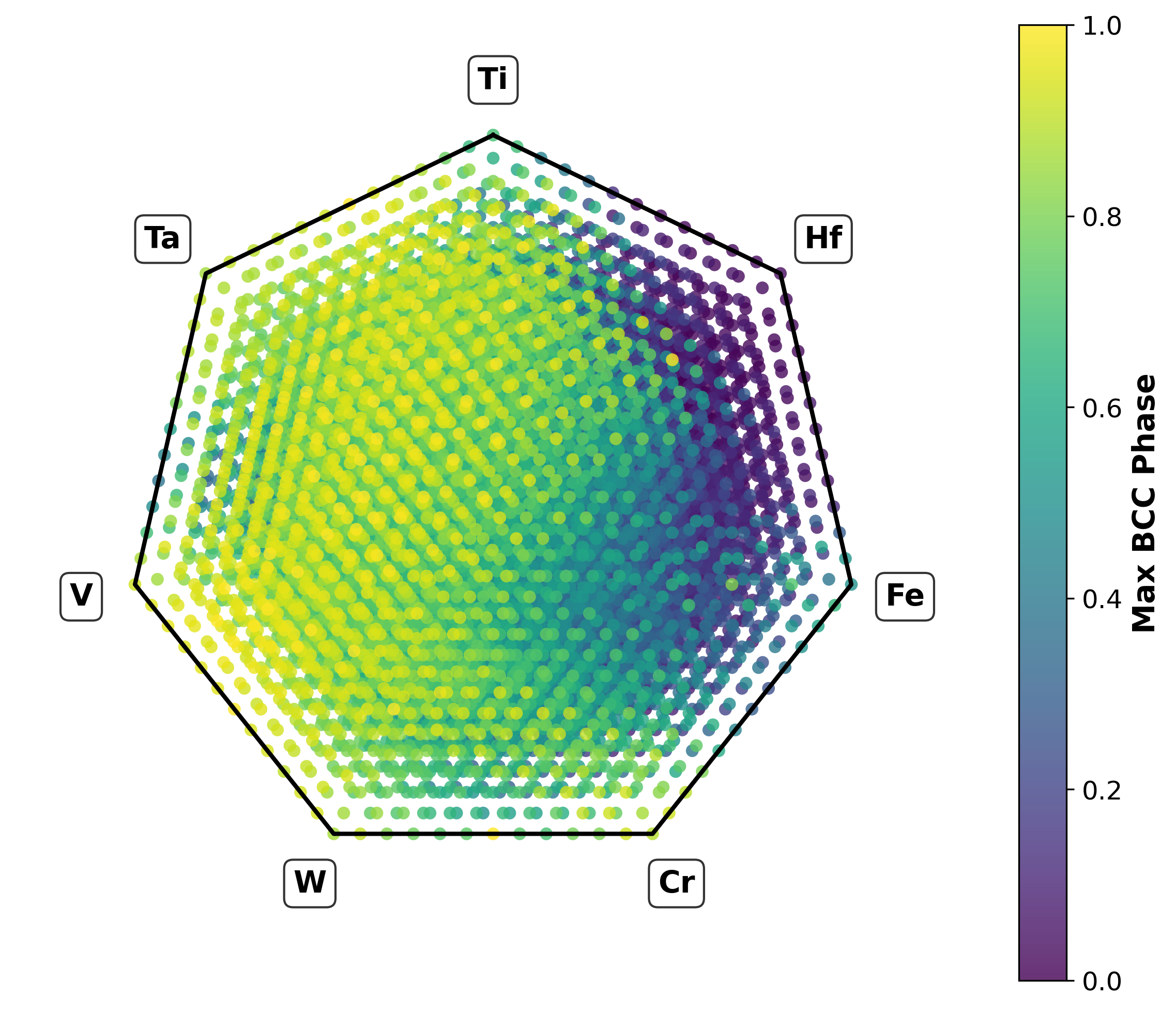}%
  }\\[1ex]
  \subfloat[Hardness]{%
    \includegraphics[width=0.32\textwidth]{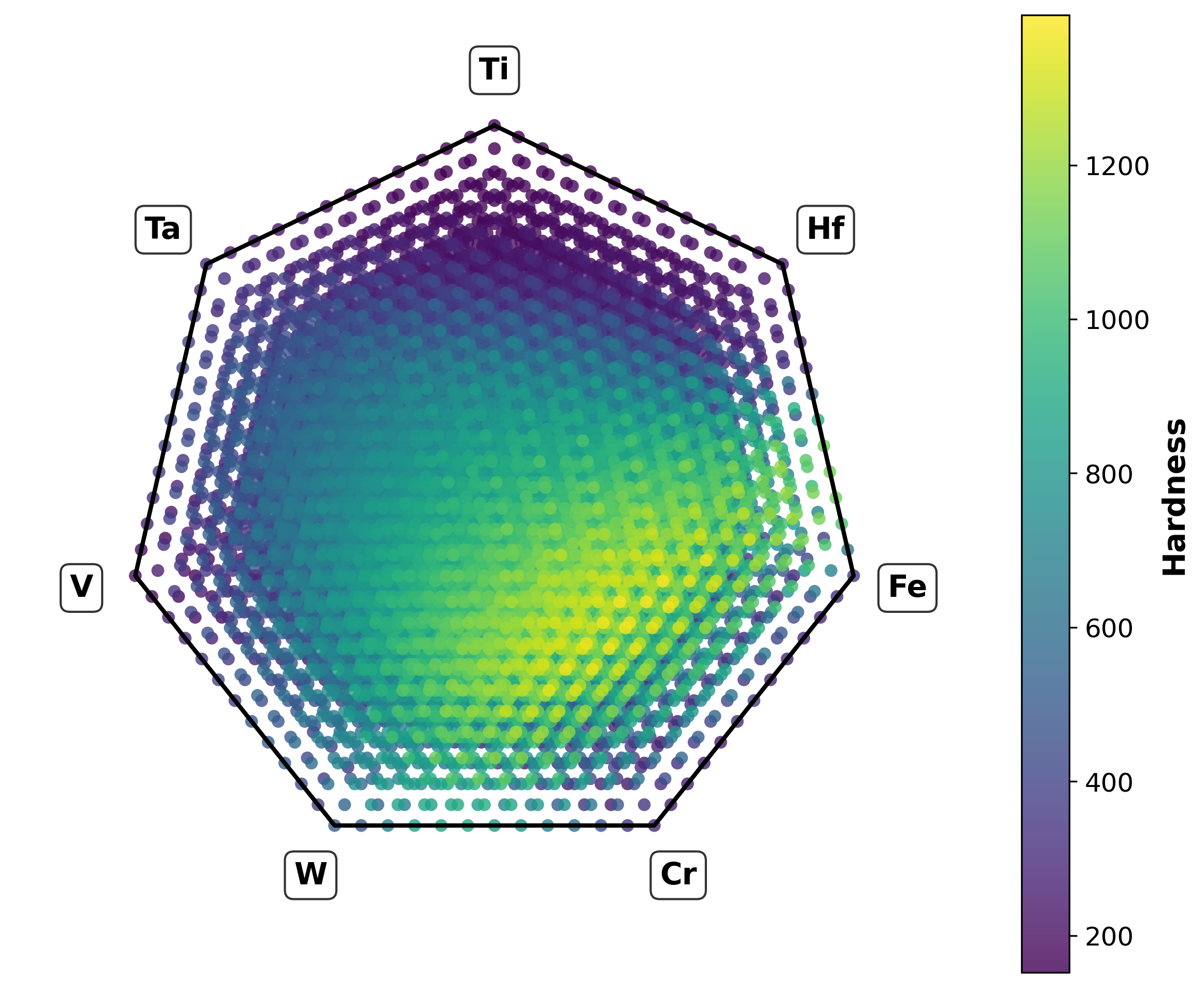}%
  }\hfill
  \subfloat[Density]{%
    \includegraphics[width=0.32\textwidth]{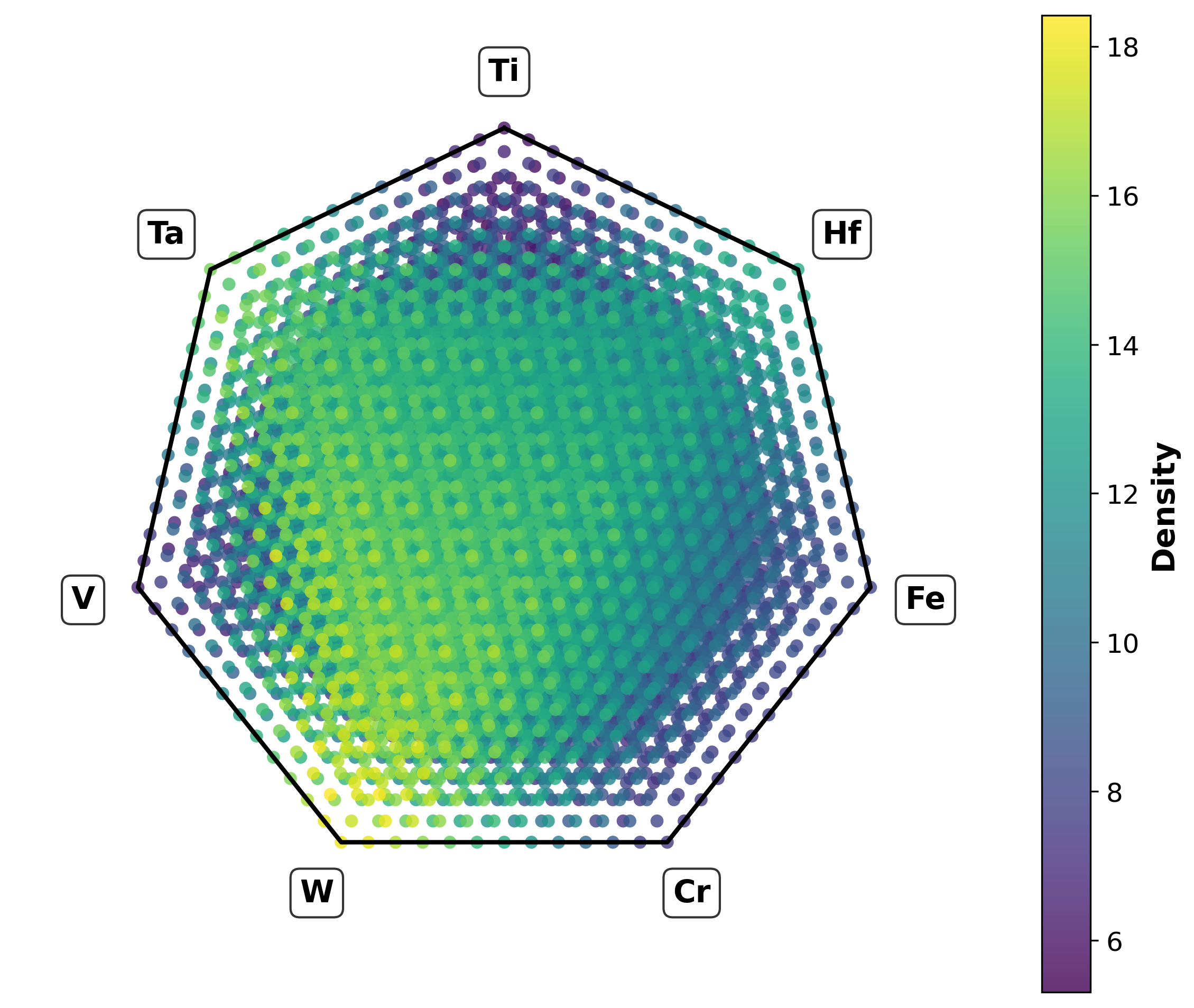}%
  }\hfill
  \subfloat[]{\phantom{\includegraphics[width=0.32\textwidth]{img/barycentric_density.png}}}%
  \caption{Barycentric composition–property maps for each of the seven alloy properties.}
  \label{fig:barycentric}
\end{figure}

Barycentric composition–property maps \cite{vela2025visualizing} in Figure~\ref{fig:barycentric} reveal how each property varies across the 7D simplex of Ti–Ta–V–W–Cr–Fe–Hf compositions.  Yield strength peaks in Cr–Fe–W–Ta–rich regions, consistent with enhanced solid solution and ordering strengthening reported for Cr–Fe‐dominant alloys \cite{vela2025visualizing}.  Thermal conductivity is highest near W‐rich corners, corroborating the low phonon scattering of W‐dominated solid solutions \cite{mora2024chemical}, while solidus temperature is maximized in W–Ta–V regions, in agreement with \emph{ab initio} melting‐point predictions for TaVCrW alloys \cite{zhou2022thermodynamics}.  The highest Pugh ratios occur in V rich compositions, indicating improved ductility for V-dominant alloys \cite{pugh1954xcii}, and Ti favors maximum BCC phase fraction–Ta–V-W mixtures, reflecting the stabilizing effect of Ti and Ta on the symmetry of the BCC. Hardness trends mirror the strength of the yield, and density distributions align with solidus patterns, underscoring the interconnected nature of mass, phase stability, and mechanical resistance in this multicomponent system \cite{singh2024alloying}.

To evaluate the effectiveness of our proposed hDGP-BO framework, we conducted a comprehensive 5-objective in-silico optimization campaign targeting the first five properties (Solidus Temperature, Thermal Conductivity, Yield Strength, BCC Phase, and Density) for maximization, while utilizing the remaining two properties (Pugh Ratio and Hardness) as auxiliary outputs to leverage inter-property correlations. The optimization was performed using KNN surrogate models trained on CALPHAD-derived data as ground truth oracles, enabling controlled comparison of different BO strategies without experimental noise.

The conventional cGP-BBO and cGP-SBO baselines employed independent single-task Gaussian processes for each property and conducted 15 iterations of isotopic queries using batch sizes of 5, selected via qEHVI acquisition. cGP-BBO and cGP-SBO respectively refers to batch BO and sequential BO using cGP.  In contrast, we implemented three variants of hierarchical deep Gaussian process models: HDGP-BO (standard DGP with doubly stochastic variational inference), HDGP-GP-BO (DGP mean function with GP variance), and HDGP-MTGP-BO (DGP mean function with multi-task GP variance). The hybrid DGP variants with GP and MTGP variance structures were motivated by literature reports indicating that DGP variance estimation through doubly stochastic inference can suffer from poor calibration and underestimation of uncertainty \cite{havasi2018inference,gnanasambandam2025deep}. Each DGP variant executed 45 total queries: every third iteration performed isotopic queries identical to cGP-BO, while the remaining iterations employed heterotopic cost-aware queries based on the task-specific cost vector $\mathbf{c} = [5.0, 10.0, 2.5, 0.5, 0.5, 0.5, 0.5]$, reflecting realistic experimental costs where thermal conductivity and solidus temperature measurements are significantly more expensive than density, hardness, or Pugh ratio evaluations.

\begin{figure}[H]
  \centering
  \includegraphics[width=0.8\textwidth]{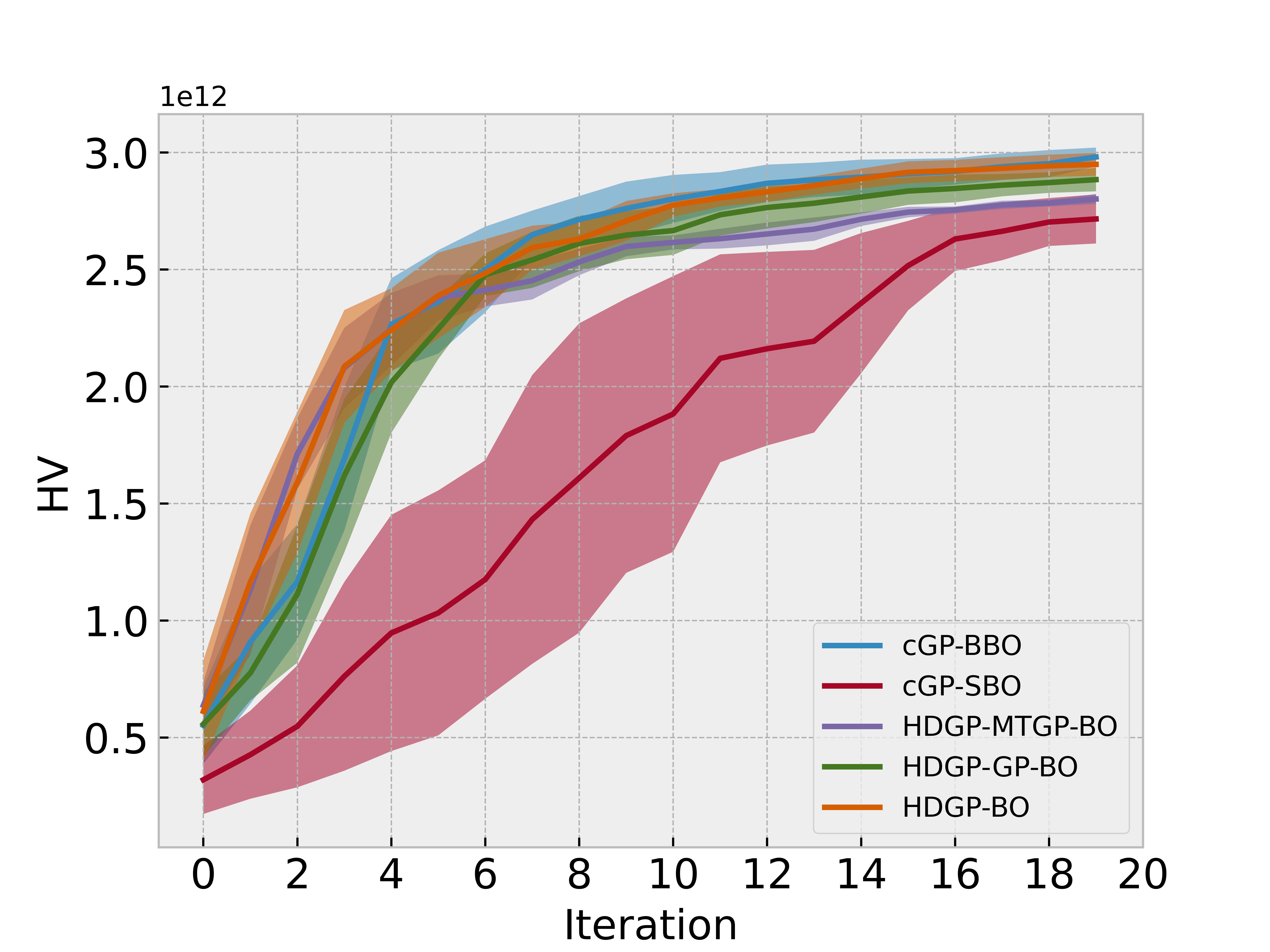}
  \caption{Hypervolume evolution across optimization iterations for cGP-BO and hDGP-BO variants. Shaded regions represent 95\% confidence intervals across multiple independent runs.}
  \label{fig:hv_comparison}
\end{figure}

Figure~\ref{fig:hv_comparison} presents the hypervolume progression throughout the optimization campaign. All methods demonstrate rapid initial improvement, reaching approximately 80\% of their final hypervolume within the first 5 iterations. Notably, the DGP-based approaches exhibit superior early-stage performance, with HDGP-GP-BO and HDGP-MTGP-BO achieving marginally higher hypervolumes during iterations 0-5. This early advantage likely stems from the DGP models' enhanced ability to capture complex composition-property relationships and exploit inter-task correlations for more informed candidate selection. However, performance differences diminish significantly in later iterations, with all methods converging to similar final hypervolumes of approximately $3.0 \times 10^{12}$. The HDGP-BO variant shows the most conservative performance, in spite of the previously mentioned limitations of doubly stochastic variance estimation in deep Gaussian processes. The cGP-SBO was used as baseline and it demonstrated inferior performance.

\begin{figure}[H]
  \centering
  \subfloat[cGP-BO Query Distribution]{%
    \includegraphics[width=0.45\textwidth]{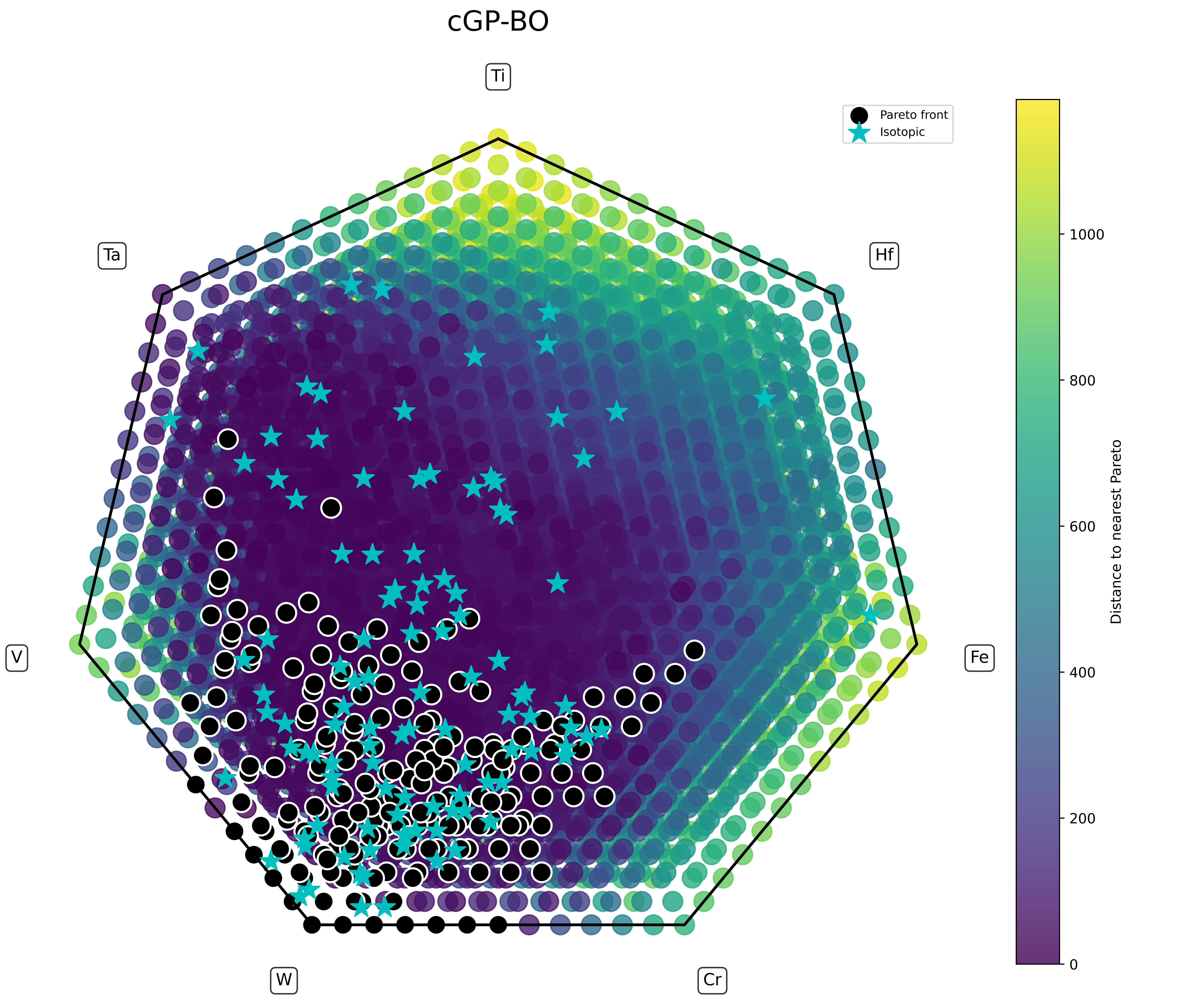}%
  }\hfill
  \subfloat[HDGP-BO Query Distribution]{%
    \includegraphics[width=0.45\textwidth]{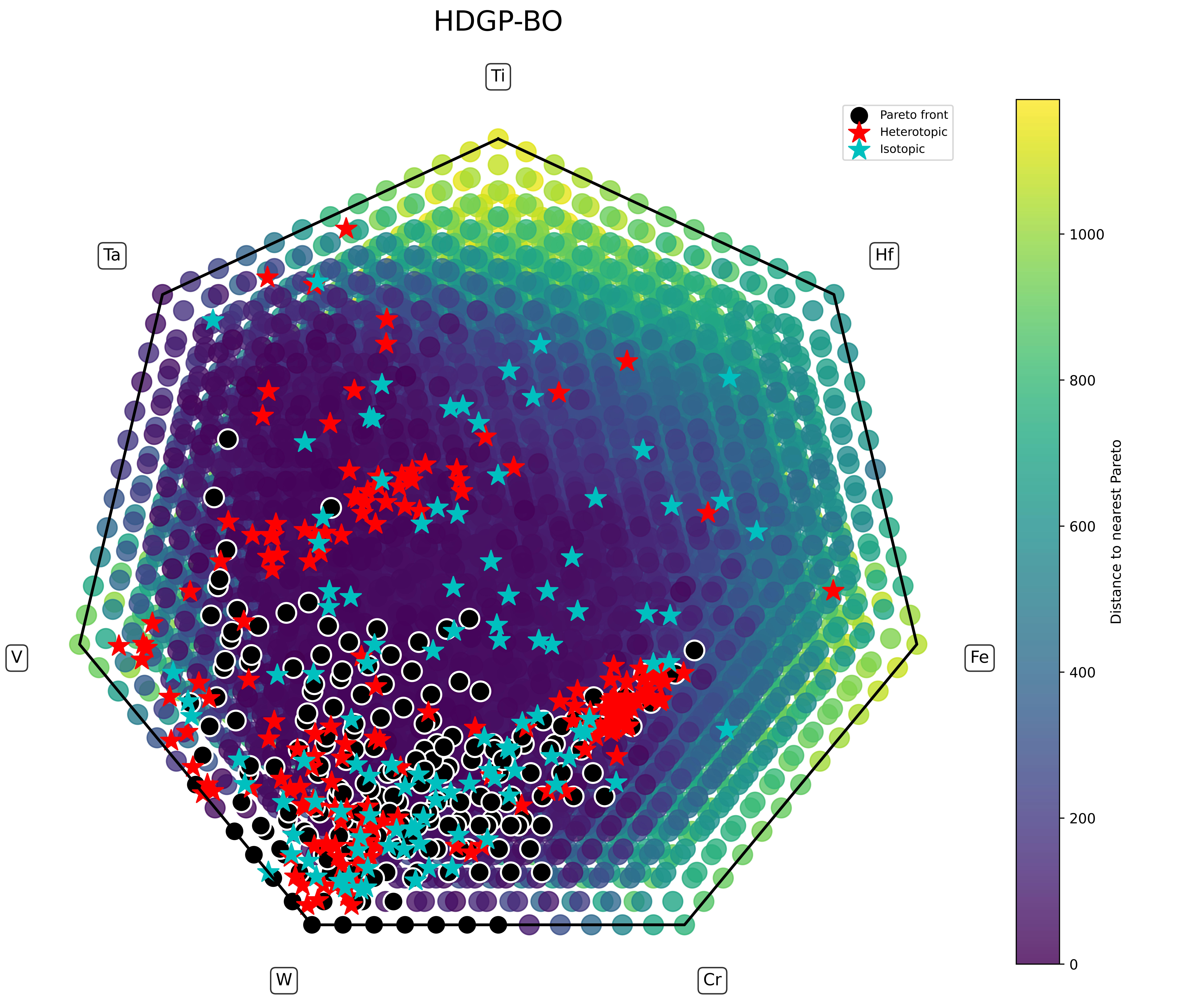}%
  }
  \caption{Barycentric composition maps showing query distributions and Pareto front evolution. Background coloring indicates distance to reference Pareto front. White circles denote final Pareto front compositions, cyan stars represent isotopic queries, and red stars indicate heterotopic queries.}
  \label{fig:query_distribution}
\end{figure}

The compositional exploration strategies reveal distinct behavioral differences between isotopic and heterotopic querying approaches (Figure~\ref{fig:query_distribution}). The cGP-BO method demonstrates concentrated exploration around high-performing regions identified early in the campaign, with isotopic queries clustering near the eventual Pareto front compositions. In contrast, HDGP-BO exhibits more diverse compositional sampling, leveraging heterotopic queries to explore regions that may be suboptimal for expensive properties but informative for cheaper auxiliary measurements. This broader exploration pattern, indicated by the wider distribution of red heterotopic query points, enables the DGP models to build more comprehensive uncertainty estimates across the entire composition space while maintaining focus on promising regions through strategic isotopic queries.

\begin{figure}[H]
  \centering
  \includegraphics[width=0.7\textwidth]{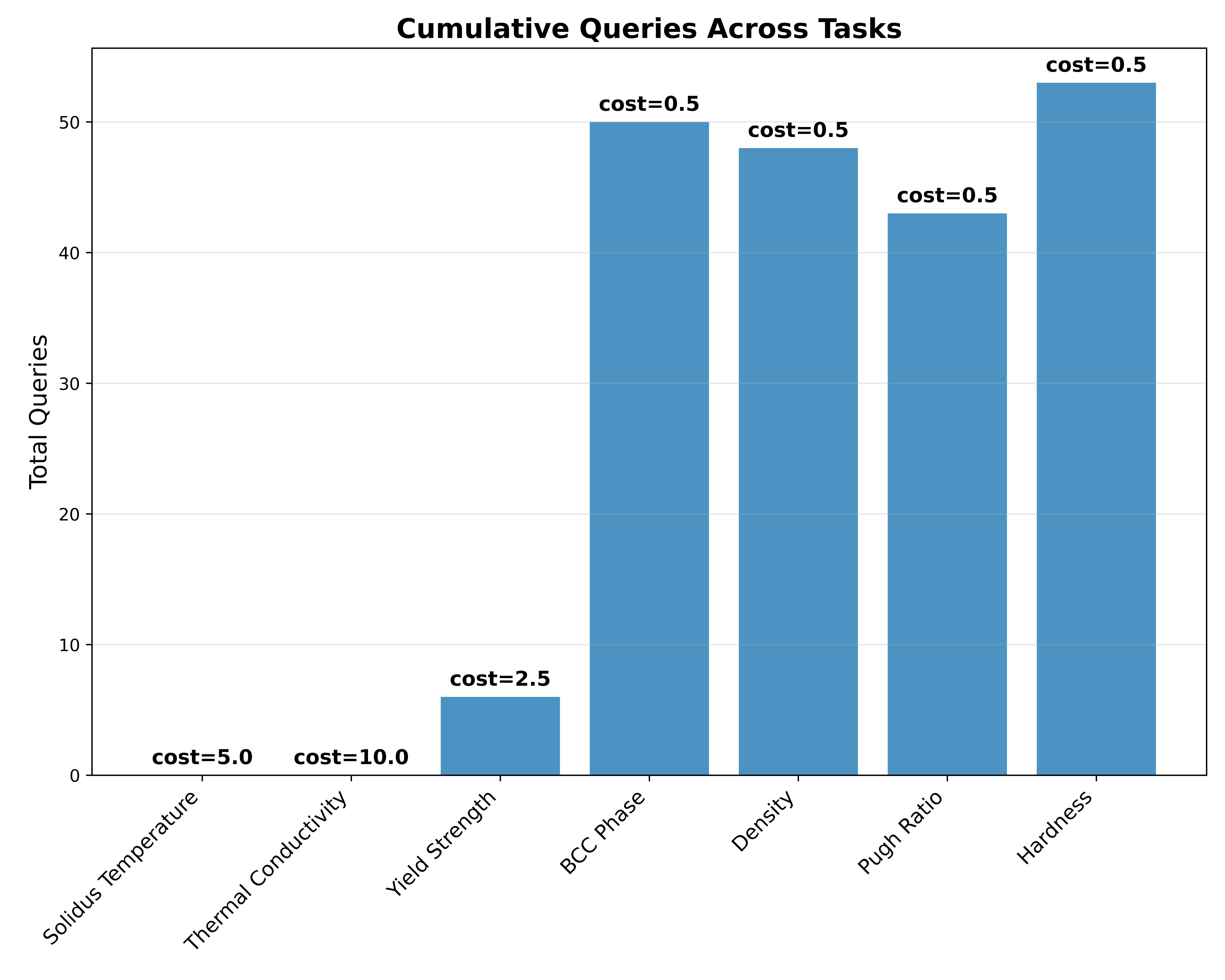}
  \caption{Cumulative query distribution across tasks, illustrating cost-aware resource allocation. Task costs are indicated above each bar.}
  \label{fig:query_allocation}
\end{figure}

The cost-aware query allocation strategy demonstrates intelligent resource management across tasks with varying evaluation costs (Figure~\ref{fig:query_allocation}). As expected, the highest-cost evaluations (Thermal Conductivity at cost=10.0 and Solidus Temperature at cost=5.0) received minimal queries, while moderate-cost Yield Strength measurements (cost=2.5) were sampled more frequently. The bulk of queries concentrated on low-cost properties (Density, Pugh Ratio, Hardness, and BCC Phase at cost=0.5 each), enabling extensive model training and uncertainty reduction for these auxiliary tasks. This heterotopic sampling strategy allows the DGP models to leverage abundant cheap data to improve predictions for expensive properties through learned inter-task correlations, exemplifying efficient budget allocation in multi-fidelity optimization scenarios.

The convergence of all methods to similar final hypervolumes, despite different exploration strategies and model complexities, suggests that the fundamental optimization landscape is well-captured by both conventional and deep GP approaches for this particular design space. While the performance differences observed in this campaign are admittedly modest, \emph{the superior early-stage performance of DGP variants indicates potential advantages in scenarios with limited evaluation budgets or when rapid initial progress is critical}. These results, combined with the more pronounced advantages demonstrated by DGPs on synthetic toy functions in the earlier sections of this work, suggest that the benefits of hierarchical modeling may be more significant in higher-dimensional materials discovery problems or design spaces with greater nonlinearity and heteroscedasticity.

Future work should explore optimization campaigns involving joint chemistry-process parameter design spaces, where the increased complexity and multi-scale interactions would likely favor the enhanced expressiveness of DGP models over conventional GPs. Such scenarios, encompassing both compositional variables and processing conditions (temperature profiles, cooling rates, mechanical working parameters), represent more realistic materials development challenges where the hierarchical feature learning capabilities of DGPs could demonstrate substantial performance gains. Additionally, experimental real-world campaigns, with their inherent measurement noise, batch effects, and non-stationary responses, may reveal more pronounced advantages for HDGP approaches compared to the relatively smooth, deterministic landscape of our CALPHAD-based in-silico evaluations. The heterotopic querying capability of DGP models provides additional flexibility for experimental campaigns where evaluation costs vary significantly across properties, enabling more sophisticated resource allocation strategies compared to traditional isotopic approaches.

\section{Conclusions}

This work presents a comprehensive investigation of heterotopic deep Gaussian process-based Bayesian optimization (HDGP-BO) for cost-aware, multi-objective materials discovery campaigns. Through systematic analysis across synthetic benchmark functions and a realistic refractory high-entropy alloy design case study, we have demonstrated the diverse performance regimes of HDGP-BO relative to conventional Gaussian process approaches.

Our synthetic function evaluations revealed three distinct performance regimes for HDGP-BO: underperformance on simple, low-dimensional problems where conventional GPs are sufficient; equivalent performance on moderately complex landscapes; and clear overperformance on highly nonlinear, multi-modal functions with complex inter-dependencies. These observations suggest that hierarchical modeling becomes advantageous when the underlying response surface exhibits structural complexity that cannot be adequately captured by shallow surrogates. As such, the relative utility of hierarchical versus conventional approaches is not universal, but depends strongly on the nature of the problem landscape.

The application to refractory high-entropy alloy optimization demonstrates the first successful implementation of cost-aware, batch HDGP-BO in an in-silico materials design setting. Our framework effectively navigated the 7-dimensional compositional space of Ti–Ta–V–W–Cr–Fe–Hf alloys, optimizing five key properties (solidus temperature, thermal conductivity, yield strength, BCC phase fraction, and density) while intelligently allocating evaluation resources based on realistic experimental cost differentials. The heterotopic querying strategy enabled efficient exploration by leveraging abundant low-cost measurements to inform decisions about expensive evaluations, exemplifying practical cost-conscious optimization in materials campaigns.

While the performance advantages of HDGP-BO were modest in our CALPHAD-based case study, the early-stage improvements observed suggest potential benefits in budget-constrained scenarios. The relatively smooth, deterministic nature of our in-silico optimization landscape may have limited the expression of HDGP advantages that become more apparent in complex, noisy experimental settings.

Future investigations should prioritize the evaluation of HDGP-BO performance in real-world experimental campaigns and expanded design spaces incorporating both chemistry and process parameters. Joint optimization of compositional variables alongside processing conditions (thermal histories, mechanical working parameters, synthesis routes) represents a significantly more challenging problem domain where the enhanced expressiveness and hierarchical feature learning capabilities of deep Gaussian processes are expected to demonstrate substantial advantages over conventional approaches. Such studies will be critical for establishing the conditions under which HDGP-BO transitions between performance regimes and for identifying the problem characteristics that most strongly favor hierarchical surrogate modeling strategies.

The cost-aware, multi-fidelity capabilities demonstrated in this work provide a foundation for next-generation materials optimization frameworks that can efficiently navigate increasingly complex design spaces while respecting realistic experimental constraints. As materials discovery campaigns evolve toward higher-dimensional, multi-scale problems, the hierarchical modeling and flexible querying strategies presented here offer promising pathways for accelerating the identification of advanced materials with tailored properties.

\section{Data Availability}
The data supporting the results of this study will be made available in a Github repository. 

\section{Code Availability}
The code supporting the results of this study will be made available in a Github repository as google colaboratory notebooks.

\section{Acknowledgements}
This material is based on work supported by the Texas A\&M University System National Laboratories Office of the Texas A\&M University System and Los Alamos National Laboratory as part of the Joint Research Collaboration Program. Any opinions, findings, conclusions or recommendations expressed in this material are those of the author(s) and do not necessarily reflect the views of the Los Alamos National Laboratory or The Texas A\&M University System. The authors acknowledge the support from the U.S. Department of Energy (DOE) ARPA-E CHADWICK Program through Project DE‐AR0001988. 
 LANL is operated by Triad National Security, LLC, for the National Nuclear Security Administration of U.S. Department of Energy (Contract No. 89233218CNA000001). SA , DA, VA and RA acknowledge BIRDSHOT Center (https://birdshot.tamu.edu), supported by the Army Research Laboratory under Cooperative Agreement (CA) Number
W911NF-22-2-0106. Calculations were carried out at Texas A\&M High-Performance Research Computing (HPRC).


\bibliographystyle{naturemag}

\bibliography{main}

\end{document}